\documentclass[11pt, reqno]{article}
\usepackage{jheppub}
\newcommand{\bea}{\begin{eqnarray}}
	\newcommand{\eea}{\end{eqnarray}}
\newcommand{\bean}{\begin{eqnarray*}}
	\newcommand{\eean}{\end{eqnarray*}}
\newcommand{\nn}{\nonumber \\}
\newcommand{\nl}{\hspace{15pt}}
\usepackage{adjustbox}
\usepackage{tikz}
\usepackage{diagbox}
\usepackage{physics}
\usepackage{booktabs}

\def\F{\mathsf{F}}
\def\L{\mathcal{L}}
\def\ii{\text{i}}

\def\ed{\ .}

\def\co{\ ,}

\def\eref#1{(\ref{#1})}

\def\Label#1{\label{#1}%
	\smash{\hbox to0pt{\raise1ex\hbox{\tiny[#1]}\hss}}}
 
\title{\boldmath Graph-Theoretic Analysis of $n$-Replica Time Evolution in the Brownian Gaussian Unitary Ensemble}

\author[a,b,c,d,e]{Tingfei Li,}
\author[d]{Jianghui Yu}
\affiliation[a]{College of Physics Science and Technology, Hebei University, Baoding 071002, China}
\affiliation[b]{Hebei Key Laboratory of High-precision Computation and Application of Quantum Field Theory, Baoding, 071002, China}
\affiliation[c]{Hebei Research Center of the Basic Discipline for Computational Physics, Baoding, 071002, China}
\affiliation[d]{Kavli Institute for Theoretical Sciences (KITS), University of Chinese Academy of Sciences, Beijing 100190, China}
\affiliation[e]{Zhejiang Institute of Modern Physics, Zhejiang University, Hangzhou, 310027, P. R. China }

\emailAdd{tfli@zju.edu.cn}
\emailAdd{yujianghui21@mails.ucas.ac.cn}

\abstract{
In this paper, we investigate the $n$-replica time evolution operator $\mathcal{U}_n(t)\equiv e^{\mathcal{L}_nt} $ for the Brownian Gaussian Unitary Ensemble (BGUE) through a graph-theoretic approach. We analyze the moments of the generator operator \( \mathcal{L}_n \), which governs the Euclidean time evolution within an auxiliary \( D^{2n} \)-dimensional Hilbert space, where $D$ represents the dimension of the Hilbert space for the original system. Explicit representations for the cases of \( n = 2 \) and \( n = 3 \) are derived, highlighting the significance of graph categorization in simplifying complex calculations. Furthermore, we discuss a general formulation applicable to arbitrary \( n \) and provide detailed examples, including the case of \( n = 4 \). Our findings reveal that the $n$-replica framework not only facilitates the evaluation of various observables but also offers insights into the interplay between Brownian disordered systems and quantum information theory.
}


\begin{document}
	\maketitle
	\flushbottom
	\section{Introduction}
	\paragraph{Motivation}
	Brownian (disordered) systems hold significant importance in various fields, including quantum gravity and quantum information. Unlike the quenched disorder found in typical disordered systems, Brownian disorder results in a local theory after taking the ensemble average, making these systems considerably easier to solve. For example, the Brownian SYK model, a crucial toy model in quantum gravity, is much more manageable than its regular counterpart \cite{Maldacena_2016,Maldacena:2016hyu}. Furthermore, the results obtained from the Brownian SYK can be utilized to evaluate the spectral form factor (SFF) of the standard SYK model \cite{Saad:2018bqo}. The connection between the Brownian SYK and de Sitter spacetime has also been explored \cite{Narovlansky:2023lfz,Milekhin:2023bjv,Stanford:2023npy}.
	
	Additionally, Brownian systems can be employed to achieve unitary $k$-designs \cite{dankert2005efficientsimulationrandomquantum,Gross_2007,ambainis2007quantumtdesignstwiseindependence,Dankert_2009}, which are distributions of unitaries that reproduce the first $k$ moments of the Haar measure. Unitary designs play a prominent role in various areas of quantum information theory \cite{PhysRevADankert,Gross2007,4262758,Scott_2008,Roy_2009,Harrow_2009,Brand_o_2016,Nakata_2017,hunterjones2019unitarydesignsstatisticalmechanics}, many-body quantum chaos \cite{Roberts_2017,Cotler_2017,jian2022lineargrowthcircuitcomplexity,fava2023designsfreeprobability}, and models of black holes \cite{Hayden_2007}.

	Recently, a model known as the Brownian Gaussian Unitary Ensemble (BGUE) was explored in \cite{Tang:2024kpv}, alongside other similar models discussed in \cite{Guo:2024zmr}. These studies consider a closed quantum system characterized by a noisy Hamiltonian
	\begin{align}
		H_{ij}(t)=\eta_{ij}(t),\eta_{ij}(t)=\eta_{ji}^*(t) \in \mathbb{C}
	\end{align}
	where $1\le i,j\le D$. $D$ is the dimension of the Hilbert space. And $\eta_{ij}(t)$ are independent Brownian Gaussian random variables with vanishing mean and non-vanishing variances
	\begin{align}
		\mathbb{E}\left(\eta_{ij}(t)\right)=0 ,~\mathbb{E}\left(\eta_{ij}(t)\eta_{kl}(t')\right) ={J\over D} \delta_{il}\delta_{jk}\delta(t-t')\ed \label{eq:GUE-variance}
	\end{align}
	Here, $\mathbb{E}$ denotes the disorder average. The model can be regarded as an infinite version of the evolution \( U = e^{-\ii H_1t_1} e^{-\ii H_2t_2} \cdots e^{-\ii H_nt_n},\ii =\sqrt{-1}\), where each \( H_i \) is independently drawn from the Gaussian Unitary Ensemble (GUE). Researchers consider this piecewise evolution and expect it to approach the Haar measure much faster than the pure evolution \( e^{-iHt} \) \cite{Cotler_2017, Roberts_2017, chen2024efficientunitarytdesignsrandom} on a timescale independent of \( D \). Similar to the Brownian SYK model, it was found that the BGUE exhibits hyperfast scrambling and the emergence of temperature \cite{lin2022infinitetemperatureshot}. In contrast to systems without disorder, when dealing with general \( n \)-replica observables in this model, we ultimately obtain an effective time evolution on \( 2n \)-contours:
	\begin{align}
		\mathcal{U}_n(t)\equiv \mathbb{E}\left(U^{\otimes n}\otimes U^{*\otimes n}\right)\equiv e^{\mathcal{L}_n t}
	\end{align}
	where $U(t) \equiv e^{-\ii H t}$, $\L_n$ is the generator operator of $\mathcal{U}_n(t)$, functioning as the ``Hamiltonian" for the Euclidean time evolution in the $D^{2n}$-dimensional Hilbert space $\mathcal{H}^{\otimes 2n}$. The calculation method bears similarities to random tensor network models \cite{Hayden_2016} and random unitary circuit models \cite{Nahum_2017,Nahum_2018,von_Keyserlingk_2018,Zhou_2020,Fisher_2023}. In \cite{Tang:2024kpv}, the author has derived explicit analytical results for one-replica ($n=1$) and two-replica ($n=2$) time evolutions, enabling the evaluation of two-point functions, OTOCs, entropy, and other quantities in the model. For instance, we find a non-decaying two-point function as $t \to \infty$ \cite{Maldacena_2003,saad2019latetimecorrelationfunctions} and non-vanishing fluctuations of the two-point function with a zero mean value \cite{stanford2020quantumnoisewormholes}. Moreover, the application of the model to $n$-design and classical shadow tomography is also discussed. As argued in the paper, many interesting observables require three (or more) replicas, such as the max-version shadow norm. This serves as one of our motivations to study the general $n$-replica case. Note that the calculation of the $n$-replica time evolution operator $\mathcal{U}_n(t)$ can be reduced to finding the general expression for the moments of $\L_n$, allowing for a straightforward construction of $\mathcal{U}_n(t)$
	\begin{align}\label{eq:Un-exp}
		\mathcal{U}_n(t)=\sum_{r=0}^{\infty}{(\L_n)^r t^r\over r!}\ed
	\end{align}
	
	
	%
	We find that the linearly independent terms, which we refer to as ``graphs,'' that appear in the moments of $\mathcal{L}_n$ are finite in number, allowing for a natural matrix representation. 
	For the case of $n=2$, author in \cite{Tang:2024kpv} previously addressed this problem by analyzing the features of the 24 individual graphs and categorizing them into 8 groups, effectively reducing the dimension of the matrix representation from 24 to 8. However, as $n$ increases, the number of individual graphs that emerge in the moments of $\mathcal{L}_n$ grows rapidly (see Table \ref{tab:Nn}), creating the challenge of how to effectively group these graphs into appropriate categories to achieve a significantly smaller-dimensional matrix representation.
	
	In this paper, we introduce a systematic notation for the graphs in the model, which facilitates a purely algebraic approach to the graph classification problem. This method enables us to derive a compact matrix representation for $\mathcal{L}_n$, making it a valuable tool for investigating the complex behaviors governed by this operator. We anticipate that the methodology presented in this paper can be applied to other models involving Brownian disorder.
	
	\paragraph{The model}
	Here, we can generalize the BGUE by introducing an intrinsic spectral term
	\begin{align}\label{eq:H}
		H_{ij}=E_{i}\delta_{ij} + \eta_{ij}(t)\ed
	\end{align}

	%
	
	%
	\paragraph{Summary of results}
	In this work, we establish a general framework to address the $n$-replica time evolution problem. We provide detailed calculations for the cases of $n=2$ and $n=3$ as illustrative examples, presenting the analytical expressions for the corresponding evolution operators $\mathcal{U}_{2}$ and $\mathcal{U}_{3}$. For $n=3$, we have 26 graph categories, so that $\L_n$ is a $26$-dimensional matrix, then we obtain  $e^{\L_3 t}= \sum_{a=1}^{26}f_a(t)e^{-\ii \mathsf{E}t}\F_a$, the analytical expressions of $f_a(t)$ are listed in Appendix \ref{appdix:u3}.

	\paragraph{Note Added}
	After completing the paper, we found an alternative approach discussed in \cite{Guo:2024zmr}. Noticing
	\begin{align}
		[\mathcal{L}_n, V^{\otimes n}\otimes V^{*\otimes n}]=0,\forall V\in U(D)\co \label{eq:GUE-symmetry}
	\end{align} 
    which allows for the determination of eigenvalues and wavefunctions for $\mathcal{L}_n$ by decomposing direct products of SU$(d)$ into irreducible representations. Moreover, these representations are constructed systematically using Young tableaux \cite{Hamermesh1962GroupTA}. Thus, they have provided a systematic method for calculating any observable in the Brownian GUE.\footnote{Similarly, one can solve Brownian GOE and Brownian GSE.} We believe that the method presented in this paper is fundamentally similar to that in \cite{Guo:2024zmr}, albeit expressed in a different form. In cases where symmetries like Eq.\eref{eq:GUE-symmetry} do not exist, the methodology discussed here can serve as a valuable toolbox.
	\paragraph{Structure of the paper}
	We first introduce the denotation for graphs in Sec. \ref{sec:graph-denotation}. Then we study the $n=2$ case as a warm-up in Sec. \ref{sec:n2}. Next, we evaluate the non-trivial $n=3$ case in Sec.  \ref{sec:n3}. Finally, we present a universal algorithm  for the general $n$ case in Sec. \ref{sec:general-n}, where the $n=4$ case is also discussed. Lastly, we provide some comments on future research in Sec. \ref{sec:discussion}.
	\section{Denotation for graphs}
	\label{sec:graph-denotation}
	In this section, we introduce a concise notation to represent arbitrary graphs, which facilitates the evaluation of graph contractions and enables a more transparent classification of graphs. We first establish the fundamental concepts, then discuss the inherent redundancies in this notation, and finally analyze the action of the group $\L_n$ on individual graphs. 
	
	At first, we introduce some notations for the graph representation of $\mathcal{U}_n$ and $\mathcal{L}_n$. We treat $\mathcal{U}_n/\mathcal{L}_n$ as a $D^{2n}$-dimensional matrix, with row and column indices labeled by two $n$-length lists: $\{I_1 J_1 I_2 J_2 \ldots I_n J_n\}$ and $\{I_1' J_1' I_2' J_2' \ldots I_n' J_n'\}$. We then adapt this to the graph representation (taking $n=2$ as an example):
	\begin{align}
		\left(\mathcal{U}_{2}(t)\right)_{I_1J_1I_2J_2;I_1'J_1'I_2'J_2'}&\equiv \mathbb{E}\left(U_{I_{1}I_{1}'}U_{J_{1}J_{1}'}^{*}U_{I_{2}I_{2}'}U_{J_{2}J_{2}'}^{*}\right)
		=\mathbb{E}\left[\adjincludegraphics[valign=c, width=0.09\textwidth]{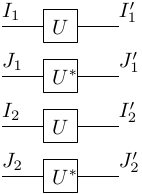}\right]=\adjincludegraphics[valign=c, width=0.09\textwidth]{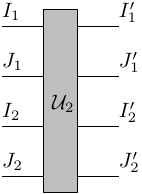}\ed
	\end{align}
    We label the indices in graphs using uppercase letters $I, J, I', J'$, where $I/I'$ represent the left/right-hand side indices of a contour $H$, and $J/J'$ denote the dual contour $H^*$. For simplicity, we label the contours (and dual contours) with lowercase letters $i, j, k, l$ (and their corresponding bars $\bar{i}, \bar{j}, \bar{k}, \bar{l}$). Therefore, we also denote
    \begin{align}
    	\mathcal{U}_{2}(t)= \adjincludegraphics[valign=c, width=0.09\textwidth]{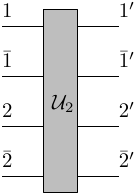}
    \end{align}
    By a similar derivation as shown in \cite{Tang:2024kpv}, it is straightforward to obtain
	\begin{align}\label{eq:Ln-def}
		\mathcal{L}_n=\mathsf{w} \mathbb{I}+{J\over D}\sum_{i\bar{j}}^n P_{i\bar{j}}-{J\over D}\sum_{ij}^n (X_{ij}+X_{\bar{i}\bar{j}})\ed
	\end{align}
	Here, \( \mathbb{I} \) denotes the identity operator, and we sometimes simply denote it as \( I \). The other two kinds of operators are represented below:
	\begin{align}
		P_{i\bar{j}}=\adjincludegraphics[valign=c, width=0.10\textwidth]{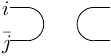},~X_{ij}=\adjincludegraphics[valign=c, width=0.10\textwidth]{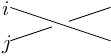}\ed
	\end{align} 
    where each line with two indices \(i,j\) at the endpoints of the graph represents a "propagator," i.e., \(\delta_{ij}\). Other trivial contours, represented by straight lines, are given by \(\delta_{I_i I_i'}\) or \(\delta_{J_i J_i'}\). Taking \(n=2\) as an example, we have
    \begin{align}
    	\left(P_{1\bar{2}}\right)_{I_1J_1I_2J_2;I_1'J_1'I_2'J_2'}&=\left[\adjincludegraphics[valign=c, width=0.11\textwidth]{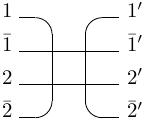}\right]_{I_1J_1I_2J_2;I_1'J_1'I_2'J_2'}=\delta_{I_1J_2}\delta_{I_1'J_2'}\delta_{J_1J_1'}\delta_{I_2I_2'}\co\nn
    	\left(X_{12}\right)_{I_1J_1I_2J_2;I_1'J_1'I_2'J_2'}&=\left[\adjincludegraphics[valign=c, width=0.11\textwidth]{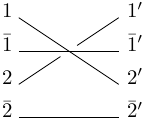}\right]_{I_1J_1I_2J_2;I_1'J_1'I_2'J_2'}=\delta_{I_{1}I_{2}'}\delta_{I_{2}I_{1}'}\delta_{J_{1}J_{1}'}\delta_{J_{2}J_{2}'}\ed
    \end{align}
	And \(\mathsf{w}\) in Eq.\ \eref{eq:Ln-def} is a factor that carries the indices of a graph
	\begin{align}
		\mathsf{w}_{I_1J_1I_2J_2\ldots I_nJ_n}=-\ii \sum_{j=1}^{n} (E_{I_j}-E_{J_j})-nJ,~\ii=\sqrt{-1}\ed
	\end{align}
	Notice that the factor \( \mathsf{w}_{I_1 J_1 I_2 J_2 \ldots I_n J_n} \) commutes with all of the terms in \(\mathcal{L}_n\), e.g.,
	\begin{align}
		\mathsf{w}_{I_{1}J_{1}I_{2}J_{2}\ldots I_{n}J_{n}}\left(P_{a\bar{b}}\right)_{I_{1}J_{1}I_{2}J_{2}\ldots I_{n}J_{n};I_{1}'J_{1}'I_{2}'J_{2}'\ldots I_{n}'J_{n}'}=\left(P_{a\bar{b}}\right)_{I_{1}J_{1}I_{2}J_{2}\ldots I_{n}J_{n};I_{1}'J_{1}'I_{2}'J_{2}'\ldots I_{n}'J_{n}'}\mathsf{w}_{I_{1}'J_{1}'I_{2}'J_{2}'\ldots I_{n}'J_{n}'}\ed
	\end{align}
    Thus, from now on, we can treat \( \mathsf{w} \) as an ordinary constant and simply write it as \( w \). For later use, we define the spectrum factor
	\begin{align}
		\mathsf{E}_{I_{1}J_{1}I_{2}J_{2}\ldots I_{n}J_{n}}=-\ii \sum_{j=1}^{n} (E_{I_j}-E_{J_j})\ed
	\end{align}
   Since \(\mathsf{w} = \mathsf{E} - nJ \mathbf{1}\), the factor \(\mathsf{E}_{I_1 J_1 I_2 J_2 \ldots I_n J_n}\) also commutes with all terms in \(\mathcal{L}_n\).
	\subsection{Notation for a graph}
	In this work, we employ the concise notation \(\mathsf{F} = \prod_{i=1}^p c_{a_i \bar{b}_i} c[\sigma]\) to represent an arbitrary graph. Here, \(0 \leq p \leq n\) denotes the number of paired vertices, and \(\sigma \equiv \sigma_1 \otimes \sigma_2 \in S_n \otimes S_n\) encodes the permutation acting on the indices appearing on the right-hand side of the graph (or equivalently, a \(D^{2n}\)-dimensional matrix \(\mathsf{F}_{I_1 J_1 I_2 J_2 \ldots I_n J_n; I_1' J_1' I_2' J_2' \ldots I_n' J_n'}\):
	\begin{align}
		\left(\prod_{i=1}^{p}c_{a_{i}\bar{b}_{i}}c[\sigma]\right)_{\{IJ\};\{I'J'\}}=\prod_{i\not\in\{a\},\bar{j}\not\in\{\bar{b}\}}\delta_{I_{i}I'_{\sigma_{1}\left(i\right)}}\delta_{J_{\bar{j}}J'_{\sigma\left(\bar{j}\right)}}\prod_{i=1}^{p}\delta_{I_{a_{i}}J_{\bar{b}_{i}}}\delta_{I'_{\sigma_{1}\left(a_{i}\right)}J'_{\sigma_{2}\left(\bar{b}_{i}\right)}}\ed
		\label{eq:Fdenotation}
	\end{align}
	%
   Here, \( c[\sigma] \) means to act the permutation on the contour labels on the right-hand side, \( c_{i \bar{j}} \) means to connect the left points \( i, \bar{j} \) and right points \( \sigma(i)', \sigma(\bar{j})' \); finally, we connect the remaining points \( i, \sigma(i)' \) and \( j, \sigma(\bar{j})' \). Taking \( c_{1 \bar{1}} c[X_{12} \otimes X_{\bar{1} \bar{2}}] = \delta_{I_1 J_1} \delta_{I_2' J_2'} \delta_{I_2 I_1'} \delta_{J_2 J_1'} \) as an example:
	\begin{equation}
		\begin{aligned}
			&\adjincludegraphics[valign=c, width=0.9\textwidth]{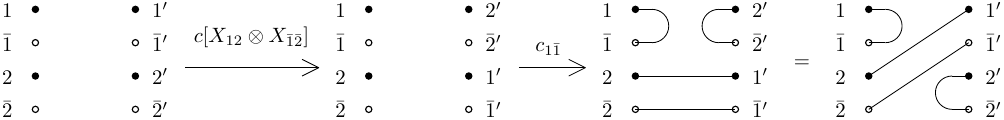}\ed
		\end{aligned} 
	\end{equation}
	%
	%
	%
	Since we use $I$ to indicate the indices of the contours $1,2$ and $J$ for $\bar{1},\bar{2}$, we do not distinguish $J_1$ and $J_{\bar{1}}$. In this paper, we use black disks to label contour vertices and circles to label the dual vertices. For simplicity, we will not draw the channel labels $i,\bar{i}$ or the indices $I_i,J_i$ in the graphs.
	
	Moreover, the denotation given in Eq.\eref{eq:Fdenotation} for the graphs makes it much easier to calculate the contraction (product) of two graphs and study the action of $\L_n$ on a certain graph.
	

	\paragraph{Symmetry of a graph}
	Before examining the action of \( P_{i \bar{j}} \) and \( X_{ij}, X_{\bar{i} \bar{j}} \) on a graph \( \mathsf{F} \), we note that the notation presented here contains some redundancy. For a graph \( \mathsf{F} = \prod_{i=1}^p c_{a_i \bar{b}_i} c[\sigma] \), the vertices for each connected pair \( (a_i, \bar{b}_i) \) are identical. To simplify matters, we define the action of a permutation on the graph
	\begin{equation}
		\begin{aligned}
			\sigma'*\F&\equiv \sigma'(\prod_{i=1}^pc_{a_i\bar{b}_i})c[\sigma\sigma'] 
			= \prod_{i=1}^pc_{\sigma'_1(a_i)\sigma'_2(\bar{b}_i)}c[\sigma\sigma']\ed
		\end{aligned}
		\label{eq:sigmaF}
	\end{equation} 
	If a permutation \( \sigma' = \sigma'_1 \otimes \sigma'_2 \) is an element of the group \( G_n(\boldsymbol{a}, \bar{\boldsymbol{b}}) \) generated by \( \{X_{a_i a_j} \otimes X_{\bar{b}_i \bar{b}_j}\}_{i,j=1,2,\ldots,p} \), then its action is trivial: $\sigma'*\F=\prod_{i=1}^pc_{a_i\bar{b_i}}c[\sigma \sigma']=\F$.
	It can be observed that \( G_n(\boldsymbol{a}, \bar{\boldsymbol{b}}) \) resembles the gauge group in quantum field theory and is isomorphic to the permutation group \( S_p \).
	
	\paragraph{Number of individual graphs}
	\begin{table}[h!]
		\centering
		\begin{tabular}{c c c}
			\toprule
			$n$ & $N_n$& $\#$ Graph Categories\\
			\midrule
			1 & 2 & 2\\
			2 & 24 & 8 \\
			3 & 720 & 26 \\
			4 & 40320& / \\
			5 & 3628800 & /\\
			\bottomrule
		\end{tabular}
		\caption{Number of individual graphs and graph categories.}
		\label{tab:Nn}
	\end{table}  
	For a given \( n \), the task is to count the number of individual graphs \( N_n \) generated by \( \mathcal{L}_n \). A simple counting yields:
	\[
	N_1 = 2, \quad N_2 = 24.
	\]
	For general \( n \), using \( |G_n(\boldsymbol{a}, \bar{\boldsymbol{b}})| = p! \) and \( |S_n \otimes S_n| = (n!)^2 \), we have
	\begin{align}
		N_n&=\sum_{\boldsymbol{a},\bar{\boldsymbol{b}}}{|S_n\otimes S_n|\over |G_n(\boldsymbol{a},\bar{\boldsymbol{b}})|}=
		\sum_{p=0}^{n}\binom{n}{p}^{2}p!\frac{\left(n!\right)^{2}}{p!}=\sum_{p=0}^{n}\left(n!\binom{n}{p}\right)^{2} \ed
		\label{eq:Nn}
	\end{align}
	One can utilize the individual graphs as a basis, then  $\mathcal{L}_n \F_i= \sum_{j=1}^{N_n}\widetilde{M}_{ji}\F_j$.
	However, as indicated in Eq.\ \eref{eq:Nn}, the dimension of \( \widetilde{M} \) grows too quickly with \( n \). A clever approach is to group these individual graphs into categories \( \F_\alpha = \sum_{i=1}^{N_{n;\alpha}} \F_{\alpha,i} \) such that
	\begin{align}\label{eq:Ln-reps}
		\mathcal{L}_n \F_\alpha = \sum_{\beta=1}^{N_{n}^{\text{ca}}} M_{\beta\alpha}F_\beta\ed 
	\end{align}
	In this paper, we further require that \( \mathsf{F}_\alpha \cap \mathsf{F}_\beta = \emptyset \) for all \( \alpha \neq \beta \). The dimension of \( M_{\beta\alpha} \), denoted as \( N_{n}^{\text{ca}} \), can be significantly smaller than \( N_n \), as illustrated in Table \ref{tab:Nn}.
	\subsection{Action of $\L_n$ on an individual graph}
	\begin{figure}[h]
		\begin{center}
			\includegraphics[width=0.8\textwidth]{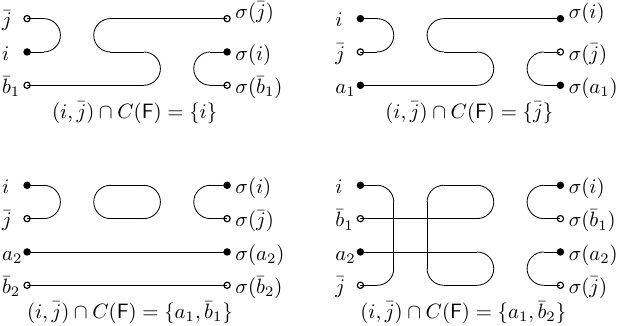}
			\caption{The four non-trivial cases of $P_{i\bar{j}}\F$. }
			\label{fig:Pijb}
		\end{center}
	\end{figure}
\begin{figure}[t]
		\begin{center}
			\includegraphics[width=0.7\textwidth]{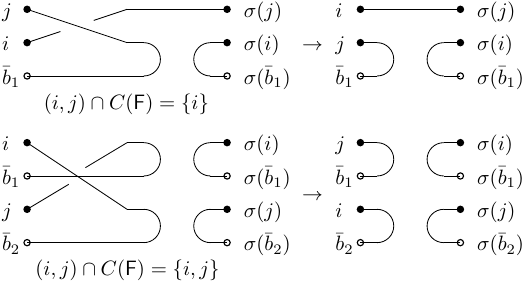}
			\caption{The two non-trivial cases of $X_{ij}\F$. }
			\label{fig:XF}
		\end{center}
	\end{figure}

	Since \( \mathcal{L}_n \) consists of \( P_{i\bar{j}} \) and \( X_{ij}, X_{\bar{i}\bar{j}} \), we first evaluate their action on an individual graph \( \mathsf{F} = \prod_{i=1}^p c_{a_i \bar{b}_i} c[\sigma] \). We denote the set of pair contours as \( C(\F) = \{(a_i, \bar{b}_i)\}_{i=1}^{p} \). For later reference, we define the set of all possible pair contours as \( \mathcal{C} \) and the exchange contours as \( \mathcal{E} \) and \( \bar{\mathcal{E}} \):
	\begin{align}
		\mathcal{C}\equiv \left\{ \left(i,\bar{j}\right)\right\}_{1\le i,\bar{j}\le n},
		\mathcal{E}\equiv \left\{ \left(i,j\right)\right\}_{1\le i< j\le n} ,\mathcal{\bar{E}}\equiv \left\{ \left(\bar{i},\bar{j}\right)\right\}_{1\le \bar{i}< \bar{j}\le n}\ed
	\end{align}
	\paragraph{Action of $P_{i\bar{j}}$}
	There are five cases:
	\begin{align}
		P_{i\bar{j}}\F=\begin{cases}
			c_{i\bar{j}}\F, & \text{if}\ (i,\bar{j})\cap C(\F)=\emptyset\\
			X_{\bar{j}\bar{b}_{1}}*\F, & \text{if}\ (i,\bar{j})\cap C(\F)=\{a_{1}\}\\
			X_{ia_{1}}*\F, & \text{if}\ (i,\bar{j})\cap C(\F)=\{\bar{b}_{1}\}\\
			D\F, & \text{if}\ (i,\bar{j})\cap C(\F)=\{a_{1},\bar{b}_{1}\}\\
			X_{a_{1}a_{2}}*\F=X_{\bar{b}_{1}\bar{b}_{2}}*\F, & \text{if}\ (i,\bar{j})\cap C(\F)=\{a_{1},\bar{b}_{2}\}
		\end{cases}\ed
	\end{align}
    Since the first case is trivial, we illustrate the other four cases in Figure \ref{fig:Pijb}.
	\paragraph{Action of $X_{ij},{X}_{\bar{i}\bar{j}}$}
	There are five cases depending on the result of \( (i,j) \cap C(\F) \). We depict two nontrivial cases in Figure \ref{fig:XF}. It is observed that we always have 
	
	\[
	X_{ij} \mathsf{F} = X_{ij} * \mathsf{F},
	\]
	and the action of \( X_{\bar{i}\bar{j}} \) is given by 
	\[
	X_{\bar{i}\bar{j}} \mathsf{F} = X_{\bar{i}\bar{j}} * \mathsf{F}.
	\]
	
	\paragraph{Action of $\L_n$} Next, we can study the action of \( \mathcal{L}_n \). Noting the condition \( \bar{u} \cap C[\F] = \emptyset \), the permutations \( X_u \) and \( X_{\bar{u}} \) act trivially on \( C(\F) \). Thus, we ultimately find a simple relation
	\begin{align}
		\mathcal{L}_{n}\F=\left(w+|C[\F]|J\right)\F+\frac{J}{D}\sum_{u\cap C[\F]=\emptyset}c_{u}\F-\frac{J}{D}\sum_{u\cap C[\F]=\emptyset}\F[\sigma X_{u}]-\frac{J}{D}\sum_{\bar{u}\cap C[\F]=\emptyset}\F[\sigma X_{\bar{u}}]\label{eq:Ln-action}
	\end{align}
	where \( |C[\F]| \) counts the number of pairs, and \( \F[\sigma \sigma'] \equiv \prod_{i=1}^p c_{a_i \bar{b}_i} c[\sigma \sigma'] \). The definitions of \( u \) and \( e \) are encoded in the summation expressions. For example, in the term $\sum_{u\cap C[\F]=\emptyset}c_{u}\F$, 
	\( u \) can be considered as a pair in \( \mathcal{C} \) that is excluded by \( C(\F) \). Meanwhile, in the term $\sum_{u\cap C[\F]=\emptyset}\F[\sigma X_{u}]$,
	\( u \) can be regarded as an element of \( \mathcal{E} \) that is excluded by \( C(\F) \). There is a commitment to taking the common set \( u \cap C[\F] \), which requires flattening \( C(\F) \) into a list of contour labels. 
	For example, taking
	 $u=\{(1,\bar{1})\},C(\F)=\{(1,\bar{2}),(2,\bar{3})\}$, then 
	\begin{align}
	   u\cap C(\F)=\{1,\bar{1}\}\cap \{1,\bar{2},2,\bar{3}\}=\{1\}\not=\emptyset\ed
	\end{align}
    We always define three graph categories, as they play a crucial role in the definition of \( \mathcal{L}_n \):
	\begin{align}
		\F_{0}\equiv c[I],\F_{1}\equiv \sum_{u} c_{u}c[I],\F_{0,X+\bar{X}}\equiv \sum_{e}c[X_{e}]+\sum_{\bar{e}}c[X_{\bar{e}}]\co
	\end{align}
	then we apply \( \mathcal{L}_n \) to the three initial categories and their descendants until a complete basis is obtained. After that, we obtain the expression for \( M \) in Eq.\ \eref{eq:Ln-reps}, which allows us to solve \( \mathcal{U}_n(t) \) using Eq.\ \eref{eq:Un-exp}. We illustrate the basic idea with examples for \( n=2 \) and \( n=3 \).

	\section{Time evolution for $n=2$}
	\label{sec:n2}
	As a warm-up, we consider \( n=2 \), which gives us four contours: \( \{1, \bar{1}, 2, \bar{2}\} \). The permutation group is \( S_2 \otimes S_2 \), where \( S_2 = \{ I, X_{12} \} \). Then
	\begin{equation}
		\begin{aligned}
			\mathcal{C}&=\left\{ \left(1,\bar{1}\right),\left(1,\bar{2}\right),\left(2,\bar{1}\right),\left(2,\bar{2}\right)\right\}\co \\
			\mathcal{E}&=\left\{ \left(1,2\right)\right\} ,\mathcal{\bar{E}}=\left\{ \left(\bar{1},\bar{2}\right)\right\} \ed
		\end{aligned}
	\end{equation}
	So we have 
	\begin{align}
		\mathcal{L}_2=w\mathbb{I}+{J\over D}\left( P_{1\bar{1}}+ P_{2\bar{2}}+ P_{1\bar{2}}+ P_{2\bar{1}}\right)-{J\over D} (X_{12}+X_{\bar{1}\bar{2}})\ed 
	\end{align}
	The number of individual graphs is \( 24 \), while the number of operators (\( P_{i\bar{j}}, X_{ij}, X_{\bar{i}\bar{j}} \)) is \( 6 \). Therefore, in principle, one needs to calculate \( 6 \times 24 = 168 \) graph contractions when evaluating \( M \) as defined earlier. Fortunately, Eq.\ \eref{eq:Ln-action} enables us to perform a simple algebraic calculation without the need to draw or recognize any graphs.
	
	\subsection{Counting graph categories}
	As discussed earlier, we aim to find the appropriate graph groups by applying \( \mathcal{L}_n \) to the three initial sets \( \mathsf{F}_0, \mathsf{F}_1, \mathsf{F}_{0,X+\bar{X}} \) and their descendants until no new graph sets are generated.
	\paragraph{$\L_2$ acts on $\F_0,\F_1,\F_{0,X+\bar{X}}$}
	\begin{itemize}
		\item $\L_2\F_0$:
		The action of $\mathcal{L}_2$ on $\F_0$ is trivial 
		\begin{align}
			\L_2\F_0 = w \F_0 + {J\over D} \F_{1} - {J\over D} \F_{0,X+\bar{X}}\ed
		\end{align}
		\item $\L_2\F_1$:
		By definition we have 
		\begin{align}
			\L_{2}\F_{1}=\left(w+J\right)\F_{1}+\frac{J}{D}\sum_{u\not=v}c_{u}c_{v}c[I]-\frac{J}{D}\left(\sum_{e\cap u=\emptyset}c_{u}c[X_{e}]+\sum_{\bar{e}\cap u=\emptyset}c_{u}c[X_{\bar{e}}]\right)\ed
		\end{align}
		Notice that the condition \( e \cap u = \emptyset \), where \( e \in \mathcal{E} \) and \( u \in \mathcal{C} \), cannot be satisfied for \( n=2 \). This is an important fact that simplifies the calculations for \( n=2 \). Therefore, we have
		\begin{align}
			\L_{2}\F_{1}=\left(w+J\right)\F_{1}+\frac{2J}{D}\F_2
		\end{align}
		where we have defined 
		\begin{align}
			\F_{2}\equiv\frac{1}{2!}\sum_{u\not=v}c_{u}c_{v}c[I]=c_{1\bar{1}}c_{2\bar{2}}c[I]+c_{1\bar{2}}c_{2\bar{1}}c[I]\ed
		\end{align}
		\item $\L_2\F_{0,X+\bar{X}}$:
		It is direct to obtain
		\begin{align}
			\L_{2}\F_{0,X+\bar{X}}&=w\F_{0,X+\bar{X}}+\frac{J}{D}\sum_{u}c_{u}\F_{0,X+\bar{X}}\nn
			&\nl-\frac{J}{D}\sum_{e,u}\left(c[X_{e}X_{u}]+c[X_{\bar{e}}X_{u}]+c[X_{e}X_{\bar{u}}]+c[X_{\bar{e}}X_{\bar{u}}]\right)\co
		\end{align}
		then we can define $\F_{1,X+\bar{X}}\equiv\sum_{u}c_{u}\F_{0,X+\bar{X}}$. 
		Notice 
		\begin{align}
			\sum_{e,u}\left(c[X_{e}X_{u}]+c[X_{\bar{e}}X_{u}]+c[X_{e}X_{\bar{u}}]+c[X_{\bar{e}}X_{\bar{u}}]\right)=2c[I]+2c[X_{12}X_{\bar{1}\bar{2}}]\ed
		\end{align}
		So we can define $\F_{0,X\bar{X}}\equiv c[X_{12}X_{\bar{1}\bar{2}}]$, then 
		\begin{align}
			\L_{2}\F_{0,X+\bar{X}}=w\F_{0,X+\bar{X}}+\frac{J}{D}\F_{1,X+\bar{X}}-\frac{2J}{D}\left(\F_{0}+\F_{0,X\bar{X}}\right)\ed
		\end{align}
	\end{itemize}
	We find that three new graph categories, \( \mathsf{F}_2, \mathsf{F}_{1,X+\bar{X}}, \mathsf{F}_{0,X\bar{X}} \), emerge when calculating \( \mathcal{L}_2^2 \). Next, we consider the action of \( \mathcal{L}_2 \) on these new categories. 
	\allowdisplaybreaks
	\paragraph{$\L_2$ acts on $\F_2,\F_{1,X+\bar{X}},\F_{0,X\bar{X}}$}
	\begin{itemize}
		\item $\L_2\F_2$:
		Notice that only the first term in Eq.\ \eref{eq:Ln-action} survives. The action of \( \mathcal{L}_2 \) on \( \F_2 \) is straightforward to obtain
		\begin{align}
			\mathcal{L}_{2}\F_{2}&=\left(w+2J\right)\F_{2}\ed
		\end{align}
		
		\item  $\L_2\F_{1,X+\bar{X}}$:
		Using Eq.\eref{eq:Ln-action}, we have 
		\begin{equation}
			\begin{aligned}
				\mathcal{L}_{2}\F_{1,X+\bar{X}}&=\left(w+J\right)\F_{1,X+\bar{X}}+\frac{J}{D}\left(\sum_{e}\sum_{g\cap u=\emptyset}c_{g}c_{u}c[X_{e}]+\left(e\leftrightarrow\bar{e}\right)\right)\co
			\end{aligned}
		\end{equation}
		so we can define 
		\begin{equation}
			\begin{aligned}
				\F_{2,X+\bar{X}}\equiv\frac{1}{4}\sum_{e}\sum_{g\cap u=\emptyset}c_{g}c_{u}c[X_{e}]+\left(e\leftrightarrow\bar{e}\right)=c_{1\bar{1}}c_{2\bar{2}}c[X_{12}]+c_{1\bar{2}}c_{2\bar{1}}c[X_{12}]\ed
			\end{aligned}
		\end{equation}
		Finally, we have 
		\begin{align}
			\mathcal{L}_{2}\F_{1,X+\bar{X}}=\left(w+J\right)\F_{1,X+\bar{X}}+{4J\over D}\F_{2,X+\bar{X}}\ed
		\end{align}
		\item  $\L_2\F_{0,X\bar{X}}$: Using Eq.\eref{eq:Ln-action}, we have 
		\begin{align}
			\mathcal{L}_{2}\F_{0,X\bar{X}}&=w\F_{0,X\bar{X}}+\frac{J}{D}\sum_{u}c_{u}c[X_{12}X_{\bar{1}\bar{2}}]-\frac{J}{D}\left(\sum_{u}c[X_{12}X_{\bar{1}\bar{2}}X_{u}]+\sum_{u}c[X_{12}X_{\bar{1}\bar{2}}X_{\bar{u}}]\right)\nn
			&=w\F_{0,X\bar{X}}+\frac{J}{D}\F_{1,X\bar{X}}-\frac{J}{D}\left(c[X_{\bar{1}\bar{2}}]+c[X_{12}]\right)\nn
			&=w\F_{0,X\bar{X}}+\frac{J}{D}\F_{1,X\bar{X}}-\frac{J}{D}\F_{0,X+\bar{X}}
		\end{align}
		where we have used $X_u=\{X_{12}\}$, so that $X_{12}X_{\bar{1}\bar{2}}X_{u}=X_{\bar{1}\bar{2}},X_{12}X_{\bar{1}\bar{2}}X_{\bar{u}}=X_{12}$.
		
	\end{itemize}
	Here, we find two new graph categories, \( \mathsf{F}_{2,X+\bar{X}} \) and \( \mathsf{F}_{1,X\bar{X}} \). We will then study the action of \( \mathcal{L}_2 \) on these categories.
	\paragraph{$\L_2$ acts on $\F_{2,X+\bar{X}},\F_{1,X\bar{X}}$}
	\begin{itemize}
		\item  $\L_2\F_{2,X+\bar{X}}$: Using Eq.\eref{eq:Ln-action}, we have 
		\begin{align}
			\mathcal{L}_{2}\F_{2,X+\bar{X}}=\left(w+2J\right)\F_{2,X+\bar{X}}\ed
		\end{align}
		\item $\L_2 \F_{1,X\bar{X}}$: Using Eq.\eref{eq:Ln-action}, we have 
		\begin{equation}
			\begin{aligned}
				\mathcal{L}_{2}\F_{1,X\bar{X}}&=\left(w+J\right)\F_{1,X\bar{X}}+\frac{J}{D}\sum_{v\cap u=\emptyset}\sum_{u}c_{v}c_{u}c[X_{12}X_{\bar{1}\bar{2}}]\\&=\left(w+J\right)\F_{1,X\bar{X}}+\frac{J}{D}\sum_{v\cap u=\emptyset}\sum_{u}c_{v}c_{u}c[I]\\&=\left(w+J\right)\F_{1,X\bar{X}}+\frac{2J}{D}\F_{2}
			\end{aligned}
		\end{equation}
		where we have used the gauge redundancy: $c_uc_vc[\sigma X_{12}X_{\bar{1}\bar{2}}]=c_uc_vc[\sigma]$. 
	\end{itemize}
	\begin{table}[t!]
		\centering
		\begin{tabular}{c c c}
			\toprule
			Rank $r$ & New Categories & $\# \F$ \\
			\midrule
			1 & $\F_0,\F_1,\F_{0,X+\bar{X}}$ & 1+4+2=7\\
			2 & $\F_2,\F_{1,X+\bar{X}},\F_{0,X\bar{X}}$ & 2+8+1=11\\
			3 & $\F_{2,X+\bar{X}},\F_{1,X\bar{X}}$ & 2+4=6\\
			4 & $\emptyset $ & 0\\
			\bottomrule
		\end{tabular}
		\caption{8 categories for $n=2$.}
		\label{tab:L2-Fs}
	\end{table}
	So far, we have found that no new categories have appeared. In summary, we list the total of 8 categories in Table \ref{tab:L2-Fs}. 
	
	\subsection{Observables}
	Comparing with the categories in \cite{Tang:2024kpv},  we can choose
	\begin{align}
		\{\F_a\}_{a=1}^8=\{\F_{0,I},\F_{1,I},\F_{0,X+\bar{X}},\F_{1,X+\bar{X}},\F_{2,I},\F_{0,X\bar{X}},\F_{2,X\bar{X}},\F_{1,X\bar{X}}\}\co
	\end{align}
	then we obtain the matrix representation of $\L_2$:
	\begin{align}
		M=\left(
		\begin{array}{cccccccc}
			w & 0 & -\frac{2 J}{D} & 0 & 0 & 0 & 0 & 0 \\
			\frac{J}{D} & J+w & 0 & 0 & 0 & 0 & 0 & 0 \\
			-\frac{J}{D} & 0 & w & 0 & 0 & -\frac{J}{D} & 0 & 0 \\
			0 & 0 & \frac{J}{D} & J+w & 0 & 0 & 0 & 0 \\
			0 & \frac{2 J}{D} & 0 & 0 & 2 J+w & 0 & 0 & \frac{2J}{D} \\
			0 & 0 & -\frac{2 J}{D} & 0 & 0 & w & 0 & 0 \\
			0 & 0 & 0 & \frac{4 J}{D} & 0 & 0 & 2 J+w & 0 \\
			0 & 0 & 0 & 0 & 0 & \frac{J}{D} & 0 & J+w \\
		\end{array}
		\right)\ed
	\end{align}
	We can use the same method as in the reference \cite{Tang:2024kpv} to solve $\mathcal{U}_2$:
	\begin{align}
		\mathcal{U}_2(t)\equiv \sum_{i=1}^{8}f_a(t)e^{-\ii \mathsf{E}t}\mathsf{F}_a\ed
	\end{align}
	where $f_a(t)$ are given by
	\begin{align}
		\begin{bmatrix}f_{1}(t)\\
			f_{2}(t)\\
			f_{3}(t)\\
			f_{4}(t)\\
			f_{5}(t)\\
			f_{6}(t)\\
			f_{7}(t)\\
			f_{8}(t)
		\end{bmatrix}=\begin{bmatrix}0 & 0 & \frac{1}{4} & \frac{1}{2} & \frac{1}{4}\\
			0 & \frac{D^{2}-2}{D\left(D^{2}-4\right)} & \frac{-1}{4(D-2)} & \frac{-1}{2D} & \frac{-1}{4(D+2)}\\
			0 & 0 & -\frac{1}{4} & 0 & \frac{1}{4}\\
			0 & \frac{-1}{D^{2}-4} & \frac{1}{4(D-2)} & 0 & \frac{-1}{4(D+2)}\\
			\frac{1}{D^{2}-1} & \frac{-2}{D^{2}-4} & \frac{1}{2(D-1)(D-2)} & 0 & \frac{1}{2(D+1)(D+2)}\\
			0 & 0 & \frac{1}{4} & -\frac{1}{2} & \frac{1}{4}\\
			\frac{-1}{D^{3}-D} & \frac{4}{D(D^{2}-4)} & \frac{-1}{2(D-1)(D-2)} & 0 & \frac{1}{2(D+1)(D+2)}\\
			0 & \frac{2}{D(D^{2}-4)} & \frac{-1}{4(D-2)} & \frac{1}{2D} & \frac{-1}{4(D+2)}
		\end{bmatrix}\times\begin{bmatrix}1\\
			e^{-Jt}\\
			e^{-(2-2D)^{-1}Jt}\\
			e^{-2Jt}\\
			e^{-(2+2D)^{-1}Jt}
		\end{bmatrix}\label{eq:fa}
	\end{align}
   Generally, we can calculate 
	\begin{align}\label{eq:4pt}
		\text{Tr}\left(O_1(t)O_2O_3(t)O_4\right)=\adjincludegraphics[valign=c, width=0.20\textwidth]{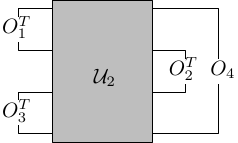}\equiv \sum_{a=1}^{8}f_a(t)\times \left(e^{-\ii\mathsf{E}t}\F_a\right)_{O_1,O_2,O_3,O_4}\ed
	\end{align}
    The noise-free case $J=0$, all $f_a(t)$ vanishes except $f_1(t)=1$, one can find 
    \begin{align}
    	\text{Tr}\left(O_1(t)O_2O_3(t)O_4\right)=\left(e^{-\ii\mathsf{E}t}\F_0\right)_{O_1,O_2,O_3,O_4}\ed
    \end{align}
    which is just the unitary time evolution for a closed system.  
	Using Eq.\eref{eq:4pt}, we can calculate the fluctuations of two-point functions, spectral form factor (SFF) and other obervables. 
	
	There is an alternative method \cite{Guo:2024zmr} to evaluate \( \mathcal{U}_2(t) \) by observing that
	\begin{align}
		[\mathcal{L}_n, V^{\otimes n}\otimes V^{*\otimes n}]=0,~\forall V\in U(D)\ed
	\end{align}
	We can always decompose a wavefunction $\psi = \psi_{i_1i_2}^{j_1j_2}\ket{i_1,j_1,i_2,j_2}_{1\bar{1}2\bar{2}}$ into two singlets $\mathbf{1}$, four adjoints $\mathbf{D}^2-\mathbf{1}$,  an antisymmetric traceless $\mathbf{A}$, two mixed $\mathbf{M}$ and a symmetric traceless $\mathbf{S}$ rank $(2,2)$ irreps of $\text{SU}(D)$. Thus, the dimension of the representation is \( 10 > 8 \), indicating that there are still some redundancies in the Young tableaux method.
	\section{Time evolution  for $n=3$}
	\label{sec:n3}
	Then we consider $n=3$, so we have six contours: $\{1,\bar{1},2,\bar{2},3,\bar{3}\}$. The permutation group is $S_3\otimes S_3$, $S_3=\{I,X_{12},X_{13},X_{23},X_{123},X_{132}\}$, then 
	\begin{equation}
		\begin{aligned}
			\mathcal{C}=\left\{ \left(i,\bar{j}\right)\right\}_{i,\bar{j}=1,2,3},
			\mathcal{E}=\left\{ \left(i,j\right)\right\}_{1\le i< j\le 3} ,\mathcal{\bar{E}}=\left\{ \left(\bar{i},\bar{j}\right)\right\}_{1\le \bar{i}< \bar{j}\le 3}\ed
		\end{aligned}
	\end{equation}
	Thus, \( |\mathcal{C}| + |\mathcal{E}| + |\mathcal{\bar{E}}| = 15 \). Unlike what we did for \( n=2 \), here we can first analyze the groups by evaluating the basic formula \eref{eq:Ln-action}. We find that the \( 720 \) individual graphs can be grouped into \( 26 \) categories. Additionally, we label them by the permutations appearing in their definitions. For \( n=3 \), the products of two different 2-cycles can yield \( X_{123} \) and \( X_{132} \), resulting in new types such as \( XX + \bar{X}\bar{X} \) and \( XX\bar{X} + \bar{X}\bar{X}X \). We present the basis in Table \ref{tab:L3-Fs}, with more explicit discussions provided in the subsection.
	\begin{table}[h!]
		\centering
		\begin{tabular}{c | c c c c c c}
			\toprule
			\diagbox{$p$}{$\#\F$}{$\F_a$}& $I$ & $X+\bar{X}$& $X\bar X$ &$XX+\bar{X}\bar{X}$ &$XX\bar{X}+\bar{X}\bar{X}X$&$XX\bar{X}\bar{X}$\\
			\midrule
			0 & 1 &6 &9&4&12&4\\
			1 & 9 &$18^{(e)},36^{(i)}$ &$9^{(ee)},36^{(ei)},36^{(ii)}$&36&$36^{(e)},72^{(i)}$&36\\
			2 & 18& $72^{(ei)},18^{(ii)}$&$36^{(ei,ei)},36^{(ei,ie)},72^{(ei,ii)}$&0&72&0\\
			3& 6 & 18&0&12&0&0\\
			\bottomrule
		\end{tabular}
		\caption{The graph categories for $n=3$.}
		\label{tab:L3-Fs}
	\end{table}

	\subsection{Grouping graphs into categories}
	Here we first define some denotations:
	\begin{align}
		X\equiv \{X_{12},X_{13},X_{23}\},~[XX]\equiv \{X_{123},X_{132}\}\ed
	\end{align}
	As discussed earlier, the number of individual graphs with \( p \) connected pairs is given by \( N_{n,p} = (n! C_n^p)^2 \). For \( n=3 \), it is
	\begin{align}
		N_{3,p}=\{36,324,324,36\}\ed
	\end{align}
    Next, we discuss the categories \( \mathsf{F}_{p,\alpha} \) for \( p = \{0, 1, 2, 3\} \). For a graph with \( p \) pairs, observe that the second term in Eq.\eref{eq:Ln-action} (\( \sum_{u \cap C[\F] = \emptyset} c_u \mathsf{F} \)) represents graphs with \( p+1 \) pairs, while the last two terms indicate applying additional 2-cycles, which are excluded by \( C(\F) \).
    \begin{itemize}
    	\item For \( p = 0 \), there are no restrictions on \( X \).
    	\item For \( p = 1 \), with \( C[\F] = c_u \), we require \( X \cap u = \emptyset \). We denote this as \( X^{(ex)}[u] \), or simply \( X^{(0)} \).
    	Similarly, we can define \( X^{(0)} = X^{(ex)}[u] \equiv \{X | X \cap u \neq \emptyset\} \). For example, if \( u = (1\bar{1}) \), then \( X^{(0)} = \{X_{23}\} \) and \( X^{(1)} = X - X^{(0)} = \{X_{12}, X_{13}\} \).
    	
    	\item For \( p > 1 \), the condition \( X \cap u = \emptyset \) cannot be satisfied when \( n = 3 \). 
    \end{itemize}
     We can generalize the definitions of \( X^{(0)} \) and \( X^{(1)} \) for a general \( p \). We define the conditions set as follows
	\begin{align}
		\mathsf{Cons}\equiv \otimes^p\{ex,in\}\ed
	\end{align}
	For simplicity, we just denote $ex=0,in=1$, and define
	\begin{align}
		\text{Con}(0,A,u)\equiv (A\cap u = \emptyset),~\text{Con}(1,A,u)\equiv (A\cap u \not= \emptyset)\co
	\end{align} 
	and for a certain connected pair $c(u_1,u_2,\ldots,u_p)$, $X^{(\alpha)}$ is a set of two-cycles:
	\begin{align}
		X^{(\alpha)}=\{X_e|\cap_{j=1}^p\text{Con}(\alpha[j],e,u_j)\}\ed
	\end{align}
	Here we take $n=3,p=2$ and $u_1=(1\bar{1}),u_2=(2\bar{2})$, then the condition set is $\textsf{Cons}=\{00,01,10,11\}$, so
	\begin{align}
		X^{(00)}=\emptyset,X^{(01)}=\{X_{23}\},X^{(10)}=\{X_{13}\},X^{(11)}=\{X_{12}\}\ed
	\end{align}
	Starting with \( \mathsf{F}_{0,I} \), we will find that \( \mathcal{L}_n^r \mathsf{F}_{0,I} \) can be expressed as a linear combination of
	\begin{align}
		\sum_{u_1,u_2,\ldots,u_p}c(u_1,u_2,\ldots,u_p)c\left[ \prod_{j=1}^{m}(X+\bar{X})^{(\alpha_j)}\right] \label{eq:blocks}
	\end{align}
	where the product in square brackets should be understood as the product between permutation sets, which we then sum over all permutations. The condition indices \( \alpha_j \) belong to the conditions set. Next, we will primarily discuss how to classify these blocks (Eq.\eref{eq:blocks}) into the appropriate graph categories.
	
	\paragraph{Zero pair: $C[\F]=0$, $\# \F=36$}
	For $p=0$, Eq.\eref{eq:blocks} becomes 
	\begin{align}
		c\left[ \prod_{j=1}^{m}(X+\bar{X})\right],m=0,1,2,3,\ldots\ed
	\end{align}
	Notice that  there is a  relation for $n=3$
	\begin{align}
		XX=3I+3[XX],~[XX]X=2X\co
	\end{align}
	so all  terms appearing in $(X+\bar{X})^m$ are
	\begin{align}
		I, X+\bar{X}, X\bar{X}, [XX]+[\bar{X}\bar{X}],[XX]\bar{X}+[\bar{X}\bar{X}]X,[XX][\bar{X}\bar{X}]\ed
	\end{align}
	Since the gauge group vanish here, there is no redundancy, then we obtain $6$ categories
	\begin{align}
		\F_{0,I},\F_{0,X+\bar{X}},\F_{0,X\bar X},\F_{0,[XX]+[\bar{X}\bar{X}]},\F_{0,[XX]\bar{X}+[\bar{X}\bar{X}]X},\F_{0,[XX][\bar{X}\bar{X}]}
	\end{align}
	where $F_{0,A}\equiv c[A]=\cup_{x\in A}c[x]$, or explicitly: 
	\begin{equation}
		\begin{aligned}
			\F_{0,I}&\equiv c[I],\F_{0,X}\equiv \sum_{e}c[X_{e}]+\sum_{\bar{e}}c[X_{\bar{e}}]\co \\
			\F_{0,X\bar{X}}&\equiv \sum_{u,\bar{e}}c[X_{\bar{e}}X_{u}],\F_{0,XX}\equiv c[X_{123}]+c[X_{132}]+c[X_{\bar{1}\bar{2}\bar{3}}]+c[X_{\bar{1}\bar{3}\bar{2}}]\co\\
			\F_{0,XX\bar{X}}&\equiv\sum_{e}c\left[X_{\bar{1}\bar{2}\bar{3}}X_{e}+X_{\bar{1}\bar{3}\bar{2}}X_{e}\right]+\sum_{\bar{e}}c\left[X_{123}X_{\bar{e}}+X_{132}X_{\bar{e}}\right]\co\\
			\F_{0,XX\bar{X}\bar{X}}&\equiv c\left[X_{123}X_{\bar{1}\bar{2}\bar{3}}\right]+c\left[X_{123}X_{\bar{1}\bar{3}\bar{2}}\right]+c\left[X_{132}X_{\bar{1}\bar{2}\bar{3}}\right]+c\left[X_{132}X_{\bar{1}\bar{3}\bar{2}}\right]\ed
		\end{aligned}
	\end{equation}
	Without introducing complexities, we neglect the dual part obtained by \( X \leftrightarrow \bar{X} \) and simply denote the categories containing \( [XX] \) and \( [\bar{X}\bar{X}] \) as \( \mathsf{F}_{0,XX} \), \( \mathsf{F}_{0,XX\bar{X}} \), and \( \mathsf{F}_{0,XX\bar{X}\bar{X}} \). 
	It is straightforward to count their sizes: \( \{1, 6, 9, 4, 12, 4\} \). Thus, we have \( 1 + 6 + 9 + 4 + 12 + 4 = 36 \), as required.
	\paragraph{One Pair: $C[\F]=1$,$\# \F=324$}
	For $p=1$, Eq.\eref{eq:blocks} becomes 
	\begin{align}
		\sum_uc_u\left[ \prod_{j=1}^{m}(X+\bar{X})^{(\alpha_j)}\right],m=0,1,2,3,\ldots, \alpha_j\in \{(0),(1)\}\ed
	\end{align}
	The size of gauge group is just $p!=1$, so there is no redundancy. For arbitrary $u=(i\bar{j})$, we find 
	\begin{align}
		\# X^{(0)}=1,~\# X^{(1)}=2,~ X=X^{(0)}+ X^{(1)}\co
	\end{align}
	so we have 
	\begin{align}
		X^{(0)}X^{(0)}=\{I\},~X^{(0)}X^{(1)}=X^{(1)}X^{(0)}=[XX],~X^{(1)}X^{(1)}=2I+2[XX]\ed
	\end{align}
	All terms which can appear are
	\begin{equation}
		\begin{aligned}
			&I,X^{(0)}+\bar{X}^{(0)},X^{(1)}+\bar{X}^{(1)},X^{(0)}\bar{X}^{(0)},X^{(0)}\bar{X}^{(1)}+\bar{X}^{(0)}X^{(1)},X^{(1)}\bar{X}^{(1)},\\
			&[XX]+[\bar{X}\bar{X}],[XX]\bar{X}^{(0)}+[\bar{X}\bar{X}]X^{(0)},[XX]\bar{X}^{(1)}+[\bar{X}\bar{X}]X^{(1)},[XX][\bar{X}\bar{X}]\ed
		\end{aligned}
	\end{equation}
	So there are $10$ categories
	\begin{align}
		\F_{1,I},\F_{1,X}^{(0)},\F_{1,X}^{(1)},\F_{1,X\bar{X}}^{(00)},\F_{1,X\bar{X}}^{(01)},\F_{1,X\bar{X}}^{(11)},\F_{1,XX},\F_{1,XX\bar{X}}^{(0)},\F_{1,XX\bar{X}}^{(1)},\F_{1,XX\bar{X}\bar{X}}
	\end{align}
	with sizes $\{1,18,36,9,36,36,36,36,72,36\}$.
	
	\paragraph{Two pairs: $C[\F]=2$,$\# \F=324$}
	For $p=2$, Eq.\eref{eq:blocks} becomes 
	\begin{align}
		\sum_{u,v}c_uc_v\left[ \prod_{j=1}^{m}(X+\bar{X})^{(\alpha_j)}\right],m=0,1,2,3,\ldots, \alpha_j\in \{(00),(01),(10),(11)\}\ed
	\end{align}
	For any two pairs $u\cap v=\emptyset$, we have 
	\begin{align}
		X^{(00)}=\emptyset, X=\{X^{(01)},X^{(10)},X^{(11)}\}\ed
	\end{align}
	So all terms appearing are  
	\begin{equation}
		\begin{aligned}
			&I,X^{(01)+(10)}+\bar{X}^{(01)+(10)},X^{(11)}+\bar{X}^{(11)},X^{(01)}\bar{X}^{(01)}+X^{(10)}\bar{X}^{(10)},\\
			&X^{(01)}\bar{X}^{(10)}+X^{(10)}\bar{X}^{(01)},X^{(01)+(10)}\bar{X}^{(11)}+X^{(11)}\bar{X}^{(01)+(10)},X^{(11)}\bar{X}^{(11)},[XX]+[\bar{X}\bar{X}],\\
			&[XX]\bar{X}^{(01)+(10)}+[\bar{X}\bar{X}]X^{(01)+(10)},[XX]\bar{X}^{(11)}+[\bar{X}\bar{X}]X^{(11)},[XX][\bar{X}\bar{X}]\ed
		\end{aligned}
	\end{equation}
	But for $p=2$, the gauge group generated by $X^{(11)}\otimes \bar{X}^{(11)}$ has size $p!=2$. We have the equivalent relation
	\begin{align}
		c_uc_vc[\sigma]=c_uc_vc\left[\sigma\cdot X^{(11)}\bar{X}^{(11)}\right]\ed
	\end{align}
	which means $X^{(11)}\bar{X}^{(11)}\sim I $, and furthermore 
	\begin{equation}
		\begin{aligned}
			[XX]&\sim[XX]X^{(11)}\bar{X}^{(11)}=X^{(10)+(01)}\bar{X}^{(11)}\co \\
			[XX][\bar{X}\bar{X}]&\sim[XX]X^{(11)}[\bar{X}\bar{X}]\bar{X}^{(11)}=X^{(10)+(01)}\bar{X}^{(10)+(01)}\co\\
			[XX]\bar{X}^{(11)}&\sim[XX]X^{(11)}\bar{X}^{(11)}\bar{X}^{(11)}=X^{(10)+(01)}\ed
		\end{aligned}
	\end{equation}
	So we only have $7$ categories
	\begin{align}
		\F_{2,I},\F_{2,X}^{(01)},\F_{2,X}^{(11)},\F_{2,X\bar{X}}^{(01,01)},\F_{2,X\bar{X}}^{(01,10)},\F_{2,X\bar{X}}^{(01,11)},\F_{2,XX\bar{X}}\ed
	\end{align}
	Their size are $\{18,72,18,36,36,72,72\}$.

	\paragraph{Three pairs: $C[\F]=3$,$\# \F=36$}
	For $p=2$, Eq.\eref{eq:blocks} becomes 
	\begin{align}
		\sum_{u_1,u_2,u_3}c_{u_1}c_{u_2}c_{u_3}\left[ \prod_{j=1}^{m}(X+\bar{X})^{(\alpha_j)}\right],m=0,1,2,3,\ldots
	\end{align}
	For three pairs $u_1,u_2,u_2$, we can only take three conditions 
	\begin{align}
		\alpha_j\in \{(011),(101),(110)\}\ed 
	\end{align}
	Any permutation \( X_e X_{\bar{e}} \) is considered a redundancy, allowing us to fix the order of \( \bar{1}, \bar{2}, \bar{3} \). This means we only need to consider permutations of \( 1, 2, 3 \). Utilizing the fact that \( X = X^{(011)} + X^{(101)} + X^{(110)} \), we identify the three building blocks: \( I, X, [XX] \).
	The respective graph categories are:
	\begin{align}
		\F_{3,I}=\cup_{u_1,u_2,u_3}c[I],\F_{3,X}=\cup_{u}c_{u_1}c_{u_2}c_{u_3}c[X],\F_{3,XX}=\cup_{u_1,u_2,u_3}c_{u_1}c_{u_2}c_{u_3}c[[XX]]\ed
	\end{align}
	They have sizes $\{6,18,12\}$.
	\subsection{Action of $\L_3$}
	Using the relations for $n=3$
	\begin{equation}
		\begin{aligned}
			XX&=3I+3[XX],~[XX]X=2X,\\
			XX_e&= X_e+[XX],~[XX]X_e=X_e[XX]=X-X_e\ed
		\end{aligned}
	\end{equation}
   and the gauge symmetry discussed earlier, it is straightforward to calculate the action of \( \mathcal{L}_3 \) on the graph categories defined in the previous section. For simplicity, we define \( \mathcal{L}_n' \) as the non-trivial action of \( \mathcal{L}_n \):
	\begin{align}
		\L_n \F_{p,\alpha}=(w+pJ)\F_{p,\alpha} + \L_n' \F_{p,\alpha}\ed
	\end{align}
	It is not hard to obtain:
	\paragraph{Zero pair: $\F_{0,I},\F_{0,X},\F_{0,X\bar X},\F_{0,XX},\F_{0,XX\bar{X}},\F_{0,XX\bar{X}\bar{X}}$}
	\allowdisplaybreaks
	\begin{align}
		\L_3' \F_{0,I}&= {J\over D}\F_{1,I}-{J\over D}\F_{0,X}\co\nn
		\L_{3}'\F_{0,X}&=\frac{J}{D}\left(\F^{(0)}_{1,X}+\F^{(1)}_{1,X}\right)-\frac{J}{D}\left(6\F_{0,I}+3\F_{0,XX}+2\F_{0,X\bar{X}}\right)\co\nn
		\L_{3}'\F_{0,X\bar{X}}&=\frac{J}{D}\left(\F_{1,X\bar{X}}^{(00)}+\F_{1,X\bar{X}}^{(11)}+\F_{1,X\bar{X}}^{(01)}\right)-\frac{J}{D}\left(3\F_{0,X}+3\F_{0,XX\bar{X}}\right)\co\nn
		\L_{3}'\F_{0,XX}&=\frac{J}{D}\F_{1,XX}-\frac{J}{D}\left(2\F_{0,X}+\F_{0,XX\bar{X}}\right)\co\nn
		\L_{3}'\F_{0,XX\bar{X}}&=\frac{J}{D}\left(\F_{1,XX\bar{X}}^{(0)}+\F_{1,XX\bar{X}}^{(1)}\right)-\frac{J}{D}\left(4\F_{0,X\bar{X}}+3\F_{0,XX}+6\F_{0,XX\bar{X}\bar{X}}\right)\co\nn
		\L_{3}'\F_{0,XX\bar{X}\bar{X}}&=\frac{J}{D}\F_{1,XX\bar{X}\bar{X}}-\frac{2J}{D}\F_{0,XX\bar{X}}\ed
	\end{align}

	\paragraph{One pair: $
		\F_{1,I},\F_{1,X}^{(0)},\F_{1,X}^{(1)},\F_{1,X\bar{X}}^{(00)},\F_{1,X\bar{X}}^{(01)},\F_{1,X\bar{X}}^{(11)},\F_{1,XX},\F_{1,XX\bar{X}}^{(0)},\F_{1,XX\bar{X}}^{(1)},\F_{1,XX\bar{X}\bar{X}}$
	}
	
	\begin{align}
		\L_{3}'\F_{1,I}&=\frac{2J}{D}\F_{2,I}-\frac{J}{D}\F_{1,X}^{(0)}\co\nn
		\L_{3}'\F_{1,X}^{(0)}&=\frac{J}{D}\F_{2,X}^{(01)}-\frac{J}{D}\left(2\F_{1,I}+2\F_{1,X\bar{X}}^{(00)}\right)\co\nn
		\L_{3}'\F_{1,X}^{(1)}&=\frac{J}{D}\left(\F_{2,X}^{(01)}+4\F_{2,X}^{(11)}\right)-
		\frac{J}{D}\left(\F_{1,XX}+\F_{1,X\bar{X}}^{(01)}\right)\co\nn
		\L_{3}'\F_{1,XX}&=\frac{2J}{D}\F_{2,X\bar{X}}^{(01,11)}-\frac{J}{D}\left(\F_{1,X}^{(1)}+\F_{1,XX\bar{X}}^{(0)}\right)\co\nn
		\L_{3}'\F_{1,X\bar{X}}^{(00)}&=\frac{J}{D}\F_{2,X\bar{X}}^{(01,01)}-\frac{J}{D}\F_{1,X}^{(0)}\co\nn
		\L_{3}'\F_{1,X\bar{X}}^{(01)}&=\frac{J}{D}\left(2\F_{2,X\bar{X}}^{(01,10)}+\F_{2,X\bar{X}}^{(01,11)}\right)-\frac{J}{D}\left(\F_{1,X}^{(1)}+\F_{1,XX\bar{X}}^{(0)}\right)\co\nn
		\L_{3}'\F_{1,X\bar{X}}^{(11)}&=\frac{J}{D}\left(\F_{2,X\bar{X}}^{(01,01)}+\F_{2,X\bar{X}}^{(01,11)}+2\F_{2,I}\right)-\frac{J}{D}\F_{1,XX\bar{X}}^{(1)}\co\nn
		\L_{3}'\F_{1,XX\bar{X}}^{(0)}&=\frac{2J}{D}\F_{2,XX\bar{X}}-\frac{J}{D}\left(\F_{1,XX}+\F_{1,X\bar{X}}^{(01)}\right)\co\nn
		\L_{3}'\F_{1,XX\bar{X}}^{(1)}&=\frac{J}{D}\left(\F_{2,XX\bar{X}}^{(01)}+2\F_{2,X+\bar{X}}^{(01)}\right)-\frac{2J}{D}\left(\F_{1,XX\bar{X}\bar{X}}+\F_{1,X\bar{X}}^{(11)}\right)\co\nn
		\L_{3}'\F_{1,XX\bar{X}\bar{X}}&=\frac{J}{D}\left(\F_{2,X\bar{X}}^{(01;01)}+2\F_{2,X\bar{X}}^{(01;10)}\right)-\frac{J}{D}\F_{1,XX\bar{X}}^{(1)}\ed
	\end{align}
	
	\allowdisplaybreaks
	\paragraph{Two pairs: $\F_{2,I},\F_{2,X}^{(01)},\F_{2,X}^{(11)},\F_{2,X\bar{X}}^{(01,01)},\F_{2,X\bar{X}}^{(01,10)},\F_{2,X\bar{X}}^{(01,11)},\F_{2,XX\bar{X}}$}
	
	\begin{align}
		\L_{3}\F_{2,I}&=\left(w+2J\right)\F_{2,I}+\frac{3J}{D}\F_{3,I}\co\nn
		\L_{3}\F_{2,X}^{(01)}&=\left(w+2J\right)\F_{2,X}^{(01)}+\frac{6J}{D}\F_{3,X}\co\nn
		\L_{3}\F_{2,X}^{(11)}&=\left(w+2J\right)\F_{2,X}^{(11)}+\frac{2J}{D}\F_{3,X}\co\nn
		\L_{3}\F_{2,X\bar{X}}^{(01,01)}&=\left(w+2J\right)\F_{2,X\bar{X}}^{(01,01)}+\frac{3J}{D}\F_{3,I}\co\nn
		\L_{3}\F_{2,X\bar{X}}^{(01,10)}&=\left(w+2J\right)\F_{2,X\bar{X}}^{(01,01)}+\frac{3J}{D}\F_{3,XX}\co\nn
		\L_{3}\F_{2,X\bar{X}}^{(01,11)}&=\left(w+2J\right)\F_{2,X\bar{X}}^{(01,11)}+\frac{3J}{D}\F_{3,XX}\co\nn
		\L_{3}\F_{2,XX\bar{X}}^{(01)}&=\left(w+2J\right)\F_{2,XX\bar{X}}^{(01)}+\frac{6J}{D}\F_{3,X}\ed
	\end{align}

	\paragraph{Three pairs: $\F_{3,I},\F_{3,X},\F_{3,XX}$}
	\begin{align}
		\L_3 \F_{3,\beta}=(w+3J)\F_{3,\beta},\beta=I,X,XX\ed
	\end{align}
	\subsection{Observables}
	We write $\mathsf{w}=-\ii\mathsf{E}-nJ$, then $M=-\ii \mathsf{E}\mathbf{1}+M_J$, $M_J$ has the explicit form
	\begin{tiny}
		\begin{align}
			\left(
			\begin{array}{cccccccccccccccccccccccccc}
				-3 J & -\frac{6 J}{D} & 0 & 0 & 0 & 0 & 0 & 0 & 0 &
				0 & 0 & 0 & 0 & 0 & 0 & 0 & 0 & 0 & 0 & 0 & 0 & 0
				& 0 & 0 & 0 & 0 \\
				-\frac{J}{D} & -3 J & -\frac{3 J}{D} & -\frac{2
					J}{D} & 0 & 0 & 0 & 0 & 0 & 0 & 0 & 0 & 0 & 0 & 0
				& 0 & 0 & 0 & 0 & 0 & 0 & 0 & 0 & 0 & 0 & 0 \\
				0 & -\frac{2 J}{D} & -3 J & 0 & -\frac{4 J}{D} & 0 &
				0 & 0 & 0 & 0 & 0 & 0 & 0 & 0 & 0 & 0 & 0 & 0 & 0
				& 0 & 0 & 0 & 0 & 0 & 0 & 0 \\
				0 & -\frac{3 J}{D} & 0 & -3 J & -\frac{3 J}{D} & 0 &
				0 & 0 & 0 & 0 & 0 & 0 & 0 & 0 & 0 & 0 & 0 & 0 & 0
				& 0 & 0 & 0 & 0 & 0 & 0 & 0 \\
				0 & 0 & -\frac{3 J}{D} & -\frac{J}{D} & -3 J &
				-\frac{2 J}{D} & 0 & 0 & 0 & 0 & 0 & 0 & 0 & 0 & 0
				& 0 & 0 & 0 & 0 & 0 & 0 & 0 & 0 & 0 & 0 & 0 \\
				0 & 0 & 0 & 0 & -\frac{6 J}{D} & -3 J & 0 & 0 & 0 &
				0 & 0 & 0 & 0 & 0 & 0 & 0 & 0 & 0 & 0 & 0 & 0 & 0
				& 0 & 0 & 0 & 0 \\
				\frac{J}{D} & 0 & 0 & 0 & 0 & 0 & -2 J & -\frac{2
					J}{D} & 0 & 0 & 0 & 0 & 0 & 0 & 0 & 0 & 0 & 0 & 0
				& 0 & 0 & 0 & 0 & 0 & 0 & 0 \\
				0 & \frac{J}{D} & 0 & 0 & 0 & 0 & -\frac{J}{D} & -2
				J & 0 & -\frac{J}{D} & 0 & 0 & 0 & 0 & 0 & 0 & 0 &
				0 & 0 & 0 & 0 & 0 & 0 & 0 & 0 & 0 \\
				0 & \frac{J}{D} & 0 & 0 & 0 & 0 & 0 & 0 & -2 J & 0 &
				-\frac{J}{D} & 0 & -\frac{J}{D} & 0 & 0 & 0 & 0 &
				0 & 0 & 0 & 0 & 0 & 0 & 0 & 0 & 0 \\
				0 & 0 & \frac{J}{D} & 0 & 0 & 0 & 0 & -\frac{2 J}{D}
				& 0 & -2 J & 0 & 0 & 0 & 0 & 0 & 0 & 0 & 0 & 0 & 0
				& 0 & 0 & 0 & 0 & 0 & 0 \\
				0 & 0 & \frac{J}{D} & 0 & 0 & 0 & 0 & 0 &
				-\frac{J}{D} & 0 & -2 J & 0 & 0 & -\frac{J}{D} & 0
				& 0 & 0 & 0 & 0 & 0 & 0 & 0 & 0 & 0 & 0 & 0 \\
				0 & 0 & \frac{J}{D} & 0 & 0 & 0 & 0 & 0 & 0 & 0 & 0
				& -2 J & 0 & 0 & -\frac{2 J}{D} & 0 & 0 & 0 & 0 &
				0 & 0 & 0 & 0 & 0 & 0 & 0 \\
				0 & 0 & 0 & \frac{J}{D} & 0 & 0 & 0 & 0 &
				-\frac{J}{D} & 0 & 0 & 0 & -2 J & -\frac{J}{D} & 0
				& 0 & 0 & 0 & 0 & 0 & 0 & 0 & 0 & 0 & 0 & 0 \\
				0 & 0 & 0 & 0 & \frac{J}{D} & 0 & 0 & 0 & 0 & 0 &
				-\frac{J}{D} & 0 & -\frac{J}{D} & -2 J & 0 & 0 & 0
				& 0 & 0 & 0 & 0 & 0 & 0 & 0 & 0 & 0 \\
				0 & 0 & 0 & 0 & \frac{J}{D} & 0 & 0 & 0 & 0 & 0 & 0
				& -\frac{J}{D} & 0 & 0 & -2 J & -\frac{J}{D} & 0 &
				0 & 0 & 0 & 0 & 0 & 0 & 0 & 0 & 0 \\
				0 & 0 & 0 & 0 & 0 & \frac{J}{D} & 0 & 0 & 0 & 0 & 0
				& 0 & 0 & 0 & -\frac{2 J}{D} & -2 J & 0 & 0 & 0 &
				0 & 0 & 0 & 0 & 0 & 0 & 0 \\
				0 & 0 & 0 & 0 & 0 & 0 & \frac{2 J}{D} & 0 & 0 & 0 &
				0 & \frac{2 J}{D} & 0 & 0 & 0 & 0 & -J & 0 & 0 & 0
				& 0 & 0 & 0 & 0 & 0 & 0 \\
				0 & 0 & 0 & 0 & 0 & 0 & 0 & \frac{J}{D} &
				\frac{J}{D} & 0 & 0 & 0 & 0 & 0 & \frac{2 J}{D} &
				0 & 0 & -J & 0 & 0 & 0 & 0 & 0 & 0 & 0 & 0 \\
				0 & 0 & 0 & 0 & 0 & 0 & 0 & 0 & \frac{4 J}{D} & 0 &
				0 & 0 & 0 & 0 & 0 & 0 & 0 & 0 & -J & 0 & 0 & 0 & 0
				& 0 & 0 & 0 \\
				0 & 0 & 0 & 0 & 0 & 0 & 0 & 0 & 0 & \frac{J}{D} & 0
				& \frac{J}{D} & 0 & 0 & 0 & \frac{2 J}{D} & 0 & 0
				& 0 & -J & 0 & 0 & 0 & 0 & 0 & 0 \\
				0 & 0 & 0 & 0 & 0 & 0 & 0 & 0 & 0 & 0 & \frac{2
					J}{D} & 0 & 0 & 0 & 0 & \frac{2 J}{D} & 0 & 0 & 0
				& 0 & -J & 0 & 0 & 0 & 0 & 0 \\
				0 & 0 & 0 & 0 & 0 & 0 & 0 & 0 & 0 & 0 & \frac{J}{D}
				& \frac{J}{D} & \frac{2 J}{D} & 0 & 0 & 0 & 0 & 0
				& 0 & 0 & 0 & -J & 0 & 0 & 0 & 0 \\
				0 & 0 & 0 & 0 & 0 & 0 & 0 & 0 & 0 & 0 & 0 & 0 & 0 &
				\frac{2 J}{D} & \frac{2 J}{D} & 0 & 0 & 0 & 0 & 0
				& 0 & 0 & -J & 0 & 0 & 0 \\
				0 & 0 & 0 & 0 & 0 & 0 & 0 & 0 & 0 & 0 & 0 & 0 & 0 &
				0 & 0 & 0 & \frac{3 J}{D} & 0 & 0 & \frac{6 J}{D}
				& 0 & 0 & 0 & 0 & 0 & 0 \\
				0 & 0 & 0 & 0 & 0 & 0 & 0 & 0 & 0 & 0 & 0 & 0 & 0 &
				0 & 0 & 0 & 0 & \frac{4 J}{D} & \frac{J}{D} & 0 &
				0 & 0 & \frac{4 J}{D} & 0 & 0 & 0 \\
				0 & 0 & 0 & 0 & 0 & 0 & 0 & 0 & 0 & 0 & 0 & 0 & 0 &
				0 & 0 & 0 & 0 & 0 & 0 & 0 & \frac{3 J}{D} &
				\frac{6 J}{D} & 0 & 0 & 0 & 0 \\
			\end{array}
			\right)\ed\nonumber
		\end{align}
	\end{tiny}
	It is direct to obtain the spectrum  of $M_J$:
	\begin{equation}
		\begin{aligned}
			\Bigg\{& 0,0,0,-\frac{3(D-2)J}{D},-\frac{3(D-1)J}{D},-\frac{2(D-1)J}{D},-\frac{2(D-1)J}{D},-\frac{2(D-1)J}{D},\\
			&-3J,-3J,-2J,-2J,-2J,-2J,-J,-J,-J,-J,-J,-J,-J,\\
			&-\frac{3(D+1)J}{D},-\frac{2(D+1)J}{D},-\frac{2(D+1)J}{D},-\frac{2(D+1)J}{D},-\frac{3(D+2)J}{D}\Bigg\}\ed
		\end{aligned}
	\end{equation}
	Then we can find the analytical expression of $\mathcal{U}_3(t)$
	\begin{align}
		e^{\L_3 t}=e^{-\ii \mathsf{E} t} \sum_{a=1}^{26}f_a(t)\F_a\ed
	\end{align}
	We plot $f_a(t)$ in Fig.\ref{fig:fa}, while the explicit expressions are given in Appendix \ref{appdix:u3}.
	\begin{figure}[h]
		\begin{center}
			\includegraphics[width=0.45\textwidth]{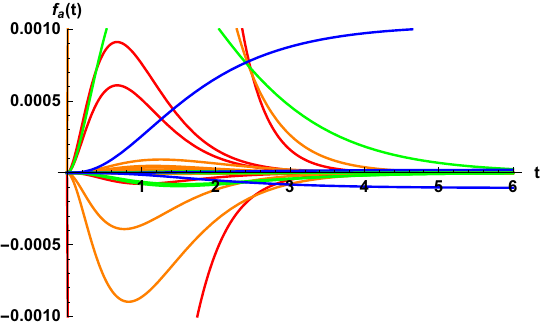}
			\includegraphics[width=0.45\textwidth]{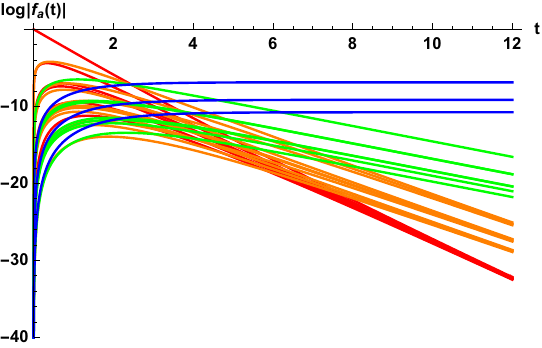}
			\caption{The pattern coefficient $f_a (t)$ or its log-plot, where we set $D = 10,J=1$. We depict the categories of $p=0,1,2,3$ with red, orange, green and blue color respectively. One can see the three categories with $p=3$ has the energy $0$, so they (blue lines) reminds finite at time $t=\infty$.   }
			\label{fig:fa}
		\end{center}
	\end{figure}
	Generally, we can calculate the observables
	\begin{align}
		\text{Tr}\left[O_{1}(t)O_{2}O_{3}(t)O_{4}O_{5}(t)O_{6}\right]=\adjincludegraphics[valign=c, width=0.20\textwidth]{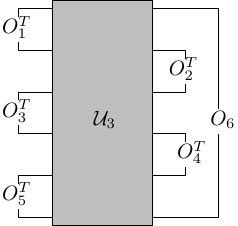}\equiv \sum_{a=1}^{26}f_a(t)\times \left(e^{-\ii \mathsf{E}t}\F_a\right)_{O_1,O_2,O_3,O_4,O_5,O_6}\label{eq:u3-obs}
	\end{align} 
	where we have used $O(t)\equiv U^\dagger O U$ and the graph representation for an operator $A$
	\begin{align}
		A_{ij}=\adjincludegraphics[valign=c, width=0.04\textwidth]{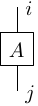}\ed
	\end{align}
	To calculate $\F_a[O_1,O_2,O_3,O_4,O_5,O_6]$, in principle, we need to contract each individual graph in the category with the operator tensor in the way as shown in Eq.\eref{eq:u3-obs}. The denotation $\F=\prod_{i=1}^pc_{a_i\bar{b}_i}c[\sigma]$ has told us the connection between points in the graph, we treat the connections on the graph as free path and the operators as bridges which connect points $i,\bar{i}$ and $\bar{i}',\overline{(i+1)}'$, here prime means the point on the right hand of the graph. So there is no need to draw the graph. We can just start with any points following the connections in the graph then finally we will return back, then we will have an operator trace
	\begin{align}
		\text{Tr}(O_iO_jO^T_k\cdots)\ed
	\end{align} 
	If there remind points we do not pass, we then do the same thing until all points in the graph are passed. The process may be a little boring, but it can be achieved by some code. Or one can just write the explicit expression of a graph with Kronecker delta function as defined in Eq.\eref{eq:Fdenotation}. 
	
	Moreover, with the expression of $\mathcal{U}_{1,2,3}(t)$, one can calculate the correlation function with general time order $\text{Tr}\left[O_{1}(t_1)O_{2}O_{3}(t_2)O_{4}O_{5}(t_3)O_{6}\right]$. For example, we consider the case $t_1<t_2<t_3$, then the ensemble-averaged time-evolution $\mathcal{U}_3(t_1,t_2,t_3)$ is
	\begin{align}
		&\mathbb{E}\left(1_D^{\otimes2}\otimes 1_D^{\otimes2}\otimes U(t_{32})\otimes U^{*}(t_{32})\right).\mathbb{E}\left(1_D\otimes 1_D\otimes U(t_{21})^{\otimes2}\otimes U^{*}(t_{21})^{\otimes2}\right).\mathbb{E}\left(U(t_{1})^{\otimes3}\otimes U^{*}(t_{1})^{\otimes3}\right)\nn
		=&\left[1_D^{\otimes2}\otimes 1_D^{\otimes2}\otimes\mathcal{U}_{1}(t_{32})\right].\left[1_D\otimes 1_D\otimes\mathcal{U}_{2}(t_{21})\right].\mathcal{U}_{3}(t_{1})\ed
	\end{align}
	where $1_D$ denotes the $D$-dimensional identity matrix. Finally, we simply need to contract \( \mathcal{U}_3(t_1, t_2, t_3) \) with the operator tensor as demonstrated in Eq.\eref{eq:u3-obs}. In this paper, we do not provide explicit expressions for the correlation functions.
	
	\section{General case}
	\label{sec:general-n}
	For general $n$, so we have $2n$ contours $\{1,\bar{1},2,\bar{2},3,\bar{3},\ldots,n,\bar{n}\}$. The permutation group is $S_n\otimes S_n$.
	As discussed before, we will meet with terms
	\begin{align}
		\sum_{u_1,u_2,\ldots,u_p}c(u_1,u_2,\ldots,u_p)c\left[ \prod_{j=1}^{m}(X+\bar{X})^{(\boldsymbol{\alpha}_j)}\right]
	\end{align}
	where recall $X^{(\boldsymbol{\alpha}_j)}$ is a set of two-cycles submitting to the condition $\boldsymbol{\alpha}_j$, i.e.,
	\begin{align}
		X^{(\boldsymbol{\alpha}_j)}\equiv \left\{X_e|\cap_{k=1}^p\text{Con}(\boldsymbol{\alpha}_j[k],e,u_k)\right\}\ed  
	\end{align}
	Since it is free to change the order of $u_1,u_2,\ldots,u_p$, there is a $S_p$ permutation redundancy for the conditions:
	\begin{align}
		\{\boldsymbol{\alpha}_j\}_{j=1,2,\ldots,p}\sim \{\sigma\boldsymbol{\alpha}_j\}_{j=1,2,\ldots,p},~\forall \sigma\in S_p\ed
	\end{align}  
	In practice, we can always write a permutation $\sigma\in S_n$ as a product of cycles $\{\sigma_{i}\}_{i=1,2,\ldots,g}$ with length $1<l_1\le l_2\ldots \le l_g$:
	\begin{align}
		X_{[l_1,l_2,\ldots,l_g]}\equiv \left\{X_{\sigma_1\sigma_2\cdots\sigma_g}||\sigma_i|=l_i,\sigma_i\cap \sigma_j=0,i\not=j \right\}\ed
	\end{align}
	Then for certain $p$, we first analyze all permutation categories  appearing  in the products $\prod_{j=1}^m X^{(\alpha_j)},m=0,1,2,\ldots$, which gives a division of the permutation group $S_n$
	\begin{align}
		S_n= \cup_{\alpha=1}^{N_X} X_{\alpha},X_{\alpha}\cap X_{\beta}=\emptyset,\forall \alpha\not=\beta\ed
	\end{align}
	So naively we have ${N_X(N_X+1)\over 2}$ graph categories:
	\begin{align}
		\F_{p,X_{\alpha}\bar{X}_{\beta}}\equiv \cup_{x\in X_{\alpha}\bar{X}_{\beta}\cup X_{\beta}\bar{X}_{\alpha} }\cup_{u_1,u_2,\ldots,u_p}c(u_1,u_2,\ldots,u_p)c[x]\ed
	\end{align} 
	Finally when $p\ge 2$, we need to consider removing the permutation redundancy of the conditions and the gauge symmetry of the graph representation, i.e,  we need to module some equivalence relation.  
	\subsection{Example: $n=4$ }
	Here we take $n=4$ as  an example, then $N_{4,p}={\{576,9216,20736,9216,576\}}$. We then define
	\begin{align}
		X_{[2]}\equiv X=\{X_{12},X_{13},X_{14},X_{23},X_{24},X_{34}\}\ed
	\end{align}
	Like what we do for $n=3$ case, we discuss the graph categories according to the number of pairs.
	\paragraph{Zero Pair $p=0,\# \F=576$}
	Due to $p=0$, there is no constricts on $X$, so we have the division of $S_4$
	\begin{align}
		S_4=I+X_{[2]}+X_{[3]}+X_{[2,2]}+X_{[4]},~\#S_4=1+6+8+3+6=24\ed
	\end{align}
	Since there is no gauge redundancy and condition permutation redundancy, we have  ${N_X(N_X+1)\over 2}=15$ graph categories:
	\begin{align}
		&\F_{0,I},\F_{0,X},\F_{0,X_{[3]}},\F_{0,X_{[2,2]}},\F_{0,X_{[4]}},\F_{0,X\bar{X}},\F_{0,X\bar{X}_{[3]}},\F_{0,X\bar{X}_{[2,2]}},\F_{0,X\bar{X}_{[4]}},\F_{0,X_{[3]}\bar{X}_{[3]}},\nn
		&\F_{0,X_{[3]}\bar{X}_{[2,2]}},
		\F_{0,X_{[3]}\bar{X}_{[4]}},\F_{0,X_{[2,2]}\bar{X}_{[2,2]}},\F_{0,X_{[2,2]}\bar{X}_{[4]}},\F_{0,X_{[4]}\bar{X}_{[4]}}\ed 
	\end{align}
	\paragraph{One Pair $p=1,\# \F = 9216$}
	For an arbitrary pair $u$, we have two conditions for $X$ and $X_{[3]}$, so that we have the division
	\begin{align}
		S_4=I+X^{(0)}+X^{(1)}+X_{[3]}^{(0)}+X_{[3]}^{(1)}+X_{[2,2]}+X_{[4]}\equiv \cup_\alpha X_{\alpha}\ed
	\end{align}
	Due to $p<2$, there is no redundancy, so we have ${7\times 8 \over 2}=28$ categories: $F_{1,X_{\alpha}\bar{X}_{\beta}}\equiv \sum_{u,x\in X_{\alpha}\bar{X}_{\beta}} c_uc[x]$.
	\paragraph{Two Pairs $p=2,\#\F =20736$}
	For any two pairs $u_1,u_2$, we find 
	\begin{align}
		X=X^{(00)}+X^{(01)}+X^{(10)}+X^{(11)},
		X_{[3]}^{(00)}=\emptyset, X_{[3]}=X_{[3]}^{(01)}+X^{(10)}_{[3]}+X_{[3]}^{(11)}\co
	\end{align}
	so we obtain the division 
	\begin{align}
		S_{4}=I+X^{(00)}+X^{(01)}+X^{(10)}+X^{(11)}+X_{[3]}^{(01)}+X_{[3]}^{(10)}+X_{[3]}^{(11)}+X_{[2,2]}+X_{[4]}\ed
	\end{align}
	Naively, we have ${10\times 11\over 2}=55$ categories. But here we need to remove the condition redundancy and gauge redundancy. For an example, changing the order of conditions, we have 
	\begin{align}
		\F_{2,X}^{(01)}=\F_{2,X}^{(10)},\F_{2,X\bar{X}}^{(00,01)}=\F_{2,X\bar{X}}^{(00,10)}\ed
	\end{align} 
	And notice the symmetry between $X$ and $\bar{X}$, we have 
	\begin{align}
		\F_{2,X\bar{X}}^{(00,01)}=\F_{2,X\bar{X}}^{(01,00)}\ed
	\end{align}
	As for the gauge redundancy, here we impose the equivalence $X^{(11)}\bar{X}^{(11)}\sim I$, so directly we have 
	\begin{align}
		\F_{2,X\bar{X}}^{(11,11)}=\F_{2,I}\ed
	\end{align}
	There are other equivalence between categories, we will not shown them here. After removing all redundancy, we will obtain the proper graph categories for $p=2$.
	\paragraph{Three pairs}
	For three pairs, we have 
	\begin{align}
		X&=X^{(001)}+X^{(010)}+X^{(100)}+X^{(011)}+X^{(101)}+X^{(110)},\nn
		X_{[3]}&=X_{[3]}^{(011)}+X_{[3]}^{(101)}+X_{[3]}^{(110)}\co
	\end{align}
    so we obtain the division 
    \begin{align}
    	S_{4}&=I+X^{(001)}+X^{(010)}+X^{(100)}+X^{(011)}+X^{(101)}+X^{(110)}\nn
    	&\nl +X_{[3]}^{(011)}+X_{[3]}^{(101)}+X_{[3]}^{(110)}+X_{[2,2]}+X_{[4]}\ed
    \end{align}
    Naively, there are ${12\times 13\over 2}=96$ categories. The condition permutation redundancy is similar to $p=2$ case and the gauge group is generated by 
	\begin{align}
		X^{(011)}\bar{X}^{(011)},X^{(101)}\bar{X}^{(101)},X^{(110)}\bar{X}^{(110)}\ed
	\end{align}
    After removing the all redundancy, the number of categories will be much smaller. 
	\paragraph{Four Pairs}
	We notice that 
	\begin{align}
		X&=X^{(0011)}+X^{(0101)}+X^{(1001)}+X^{(0110)}+X^{(1010)}+X^{(1100)}\co\nn
		X_{[3]}&=X_{[3]}^{(0111)}+X_{[3]}^{(1011)}+X_{[3]}^{(1101)}+X_{[3]}^{(1110)}\co
	\end{align}
	which gives the division 
	\begin{align}
		S_{4}&=I+X^{(0011)}+X^{(0101)}+X^{(1001)}+X^{(0110)}+X^{(1010)}+X^{(1100)}\nn
		&\nl +X_{[3]}^{(011)}+X_{[3]}^{(101)}+X_{[3]}^{(110)}+X_{[2,2]}+X_{[4]}\ed
	\end{align}
	Notice the gauge group is generated by any permutation like $X_e\bar{X}_e,\forall e$.
	So we can fix the order of $\bar{i}$ to be $\bar{1}\bar{2}\bar{3}\bar{4}$, then we just need to consider the categories
	\begin{align}
		\F_{\alpha}\equiv \cup_{u_1,u_2,u_3,u_4}c(u_1,u_2,u_3,u_4)c[X_\alpha]\ed
	\end{align}
	Notice the permutation redundancy of the condition, we have 
	\begin{align}
		X^{(0011)}\sim X^{(0101)}\sim X^{(1001)}\sim X^{(0110)}\sim X^{(1010)}\sim X^{(1100)},X_{[3]}^{(011)}\sim X_{[3]}^{(101)}\sim X_{[3]}^{(110)}\co
	\end{align} 
	so we only have five categories
	\begin{align}
		\F_{4,I},\F_{4,X},\F_{4,X_{[3]}},\F_{4,X_{[2,2]}},\F_{4,X_{[4]}}\ed
	\end{align}
	\section{Discussion and Outlook}
	\label{sec:discussion}

	The graph-based framework developed in this work provides a powerful approach for analyzing the $n$-replica physics of the BGUE model. The explicit constructions for the $n=2$ and $n=3$ cases demonstrate how the increasing complexity of the graphs can be managed by organizing them into distinct categories. The general formulation outlined in Section \ref{sec:general-n} suggests that this approach can be extended to arbitrary $n$, though the proliferation of graph structures poses a significant challenge. Developing efficient methods for packing and manipulating the graphs will be an important direction for future research.
	
	A key insight is that the factor $w$, which encodes the dependence on the system spectrum, commutes with the graph-based operators. This allows for the analysis of the system with infinite-temperature GUE noise without introducing additional complexity.
	
	The techniques discussed in this paper can be applied to other models, such as the Brownian Gaussian Orthogonal Ensemble (BGOE) and Brownian Gaussian Symplectic Ensemble (BGSE). For BGOE, we have 
	
	\begin{align}
		\L_n=\frac{-JD}{D+1}\left[n\mathbb{I}+\sum_{i<j}\left(P_{ij}+P_{\bar{i}\bar{j}}\right)-\sum_{i,\bar{j}}P_{i\bar{j}}\right]+\frac{-J}{D+1}\left[n+\sum_{i<j}X_{ij}+X_{\bar{i}\bar{j}}-\sum_{i\bar{j}}X_{i\bar{j}}\right]\co
	\end{align}
	and for BGSE:
	\begin{align}
		\L_n=\frac{-J}{D-1}\left[2n(D-1)\mathbb{I}-D\sum_{i<j}\left(P_{ij}+P_{\bar{i}\bar{j}}\right)-2D\sum_{i,\bar{j}}P_{i\bar{j}}+4\sum_{i<j}\left(X_{ij}+X_{\bar{i}\bar{j}}\right)+2\sum_{i,\bar{j}}X_{i\bar{j}}\right]\ed
	\end{align}
	We find there are new operators $P_{ij},P_{\bar{i}\bar{j}},X_{i\bar{j}}$, so an individual graph appearing in $\L_n^r$ can be written as 
	\begin{align}
		\F=\prod_{i=1}^{p_1} c_{a_i a'_i}\prod_{i=1}^{p_2} c_{\bar{b}_i \bar{b}'_i}\prod_{i=1}^{p_3}c_{r_i\bar{s}_i}c\left[\sigma\right]\ed
	\end{align}
	%
	So there will be more individual graphs, which can be challenging to organize into appropriate categories. While this task may seem intricate, it is fundamentally a straightforward endeavor. However, as discussed in the introduction, the method outlined in this paper is essentially equivalent to the approach detailed in \cite{Guo:2024zmr}. The inherent symmetry of $\mathcal{L}_n$ for BGUE, BGOE and BGSE simplifies its diagonalization, a well-studied problem in the field of mathematics: the decomposition of the direct product of the fundamental representations. We therefore anticipate that the techniques employed in this paper can be successfully applied to models lacking obvious symmetry, thereby highlighting the power and versatility of the proposed method.
	\section*{Acknowledgment}
	The author would like to thank Haifeng Tang, Cheng Peng, Yingyu Yang, and Yanyuan Li for their valuable discussions and comments during the preparation of this work.
	\appendix
	\section{Expression of $\mathcal{U}_3(t)$}
	\label{appdix:u3}
	As discussed in Sec.\ref{sec:n3}, we write $\mathcal{U}_3=\sum_{a=1}^{26}f_a(t)e^{-\ii \mathsf{E}t}\F_a$. Here we list the explicit expressions for $f_a$'s. For simplicity, we denote 
	\begin{align}
		f_a(t)=\boldsymbol{v}_a\cdot \boldsymbol{\xi}(t)
	\end{align}
	where $\boldsymbol{\xi}(t)$ are $10$ wavefunctions. 
	\begin{align}
		\boldsymbol{\xi}(t)=\left\{e^{-J t},e^{-2 J t},e^{-\frac{2 (D+1) J t}{D}},e^{-\frac{2 (D-1) J t}{D}},e^{-3 J t},e^{-\frac{3
				(D+1) J t}{D}},e^{-\frac{3 (D-1) J t}{D}},e^{-\frac{3 (D+2) J t}{D}},e^{-\frac{3 (D-2) J
				t}{D}},1\right\}
	\end{align}
	and the $26$ vectors are 
	\allowdisplaybreaks
	\begin{align}
		\boldsymbol{v}_1&=\left\{0,0,0,0,\frac{1}{2},\frac{2}{9},\frac{2}{9},\frac{1}{36},\frac{1}{36},0\right\},\nn
		\boldsymbol{v}_2&=\left\{0,0,0,0,0,\frac{1}{9},-\frac{1}{9},\frac{1}{36},-\frac{1}{36},0\right\},\nn
		\boldsymbol{v}_3&=\left\{0,0,0,0,-\frac{1}{18},0,0,\frac{1}{36},\frac{1}{36},0\right\},\nn
		\boldsymbol{v}_4&=\left\{0,0,0,0,-\frac{1}{6},\frac{1}{18},\frac{1}{18},\frac{1}{36},\frac{1}{36},0\right\},\nn
		\boldsymbol{v}_5&=\left\{0,0,0,0,0,-\frac{1}{18},\frac{1}{18},\frac{1}{36},-\frac{1}{36},0\right\},\nn
		\boldsymbol{v}_6&=\left\{0,0,0,0,\frac{1}{6},-\frac{1}{9},-\frac{1}{9},\frac{1}{36},\frac{1}{36},0\right\},\nn
		\boldsymbol{v}_7&=\Bigg\{0,\frac{5-D^2}{18 D-2 D^3},\frac{D^2+3 D-2}{4 (D-2) (D+1) (D+4)},\frac{D^2-3 D-2}{4 (D-4) (D-1)
			(D+2)},\nn
		&\nl-\frac{20-9 D^2}{72 D-18 D^3},-\frac{2 (D+2)}{9 (D+1) (D+3)},-\frac{2 (D-2)}{9 \left(D^2-4
			D+3\right)},-\frac{1}{36 (D+4)},\frac{1}{144-36 D},0\Bigg\},\nn
		\boldsymbol{v}_8&=\Bigg\{0,0,\frac{D^2+3 D-2}{4 (D-2) (D+1) (D+4)},\frac{-D^2+3 D+2}{4 \left(D^3-3 D^2-6
			D+8\right)},\frac{4}{36-9 D^2},\nn
		&\nl-\frac{1}{9 D+9},\frac{1}{9 (D-1)},-\frac{1}{36 (D+4)},\frac{1}{36
			(D-4)},0\Bigg\},\nn
		\boldsymbol{v}_9&=\Bigg\{0,\frac{1}{18-2 D^2},-\frac{D+2}{4 \left(D^3+3 D^2-6 D-8\right)},-\frac{D-2}{4 (D-4) (D-1)
			(D+2)},\nn
		&\nl \frac{2}{9 \left(D^2-4\right)},-\frac{2 D+3}{18 (D+1) (D+3)},\frac{2 D-3}{18 (D-3)
			(D-1)},-\frac{1}{36 (D+4)},\frac{1}{36 (D-4)},0\Bigg\},\nn
		\boldsymbol{v}_{10}&=\Bigg\{0,-\frac{D^2-5}{2 D \left(D^2-9\right)},\frac{D^2+3 D-2}{4 (D-2) (D+1) (D+4)},\frac{D^2-3 D-2}{4
			(D-4) (D-1) (D+2)},\frac{20-D^2}{72 D-18 D^3},\nn
		&\nl-\frac{2}{9 (D+1) (D+3)}, \frac{2}{9 \left(D^2-4
			D+3\right)},-\frac{1}{36 (D+4)},\frac{1}{144-36 D},0\Bigg\},\nn
		\boldsymbol{v}_{11}&=\Bigg\{0,\frac{1}{18 D-2 D^3},-\frac{D+2}{4 \left(D^3+3 D^2-6 D-8\right)},\frac{D-2}{4 (D-4) (D-1)
			(D+2)},-\frac{D^2+4}{72 D-18 D^3},\nn
		&\nl -\frac{1}{18 (D+1) (D+3)},\frac{1}{18 \left(D^2-4
			D+3\right)},-\frac{1}{36 (D+4)},\frac{1}{144-36 D},0\Bigg\},\nn
		\boldsymbol{v}_{12}&=\Bigg\{0,\frac{1}{D \left(D^2-9\right)},\frac{1}{2 (D-2) (D+1) (D+4)},\frac{1}{2 \left(D^3-3 D^2-6
			D+8\right)},\frac{8-D^2}{72 D-18 D^3},\nn
		&\nl \frac{1}{9 \left(D^2+4 D+3\right)},-\frac{1}{9 \left(D^2-4
			D+3\right)},-\frac{1}{36 (D+4)},\frac{1}{144-36 D},0\Bigg\},\nn
		\boldsymbol{v}_{13}&=\Bigg\{0,-\frac{1}{18 D-2 D^3},-\frac{D+2}{4 \left(D^3+3 D^2-6 D-8\right)},\frac{D-2}{4 (D-4) (D-1)
			(D+2)},\frac{4-3 D^2}{72 D-18 D^3},\nn
		&\nl -\frac{D+2}{18 (D+1) (D+3)},-\frac{D-2}{18 \left(D^2-4
			D+3\right)},-\frac{1}{36 (D+4)},\frac{1}{144-36 D},0\Bigg\},\nn
		\boldsymbol{v}_{14}&=\Bigg\{0,\frac{1}{2 \left(D^2-9\right)},-\frac{D+2}{4 \left(D^3+3 D^2-6 D-8\right)},-\frac{D-2}{4 (D-4)
			(D-1) (D+2)},\frac{2}{9 \left(D^2-4\right)},\nn
		&\nl \frac{D}{18 \left(D^2+4 D+3\right)},-\frac{D}{18
			\left(D^2-4 D+3\right)},-\frac{1}{36 (D+4)},\frac{1}{36 (D-4)},0\Bigg\},\nn
		\boldsymbol{v}_{15}&=\Bigg\{0,0,\frac{1}{2 (D-2) (D+1) (D+4)},-\frac{1}{2 \left(D^3-3 D^2-6 D+8\right)},\frac{1}{36-9
			D^2},\frac{1}{18 D+18},\nn
		&\nl \frac{1}{18-18 D},-\frac{1}{36 (D+4)},\frac{1}{36 (D-4)},0\Bigg\},\nn
		\boldsymbol{v}_{16}&=\Bigg\{0,\frac{1}{9 D-D^3},\frac{1}{2 (D-2) (D+1) (D+4)},\frac{1}{2 \left(D^3-3 D^2-6
			D+8\right)},-\frac{8-3 D^2}{72 D-18 D^3},\nn
		& \nl \frac{D+2}{9 \left(D^2+4 D+3\right)},\frac{D-2}{9 (D-3)
			(D-1)},-\frac{1}{36 (D+4)},\frac{1}{144-36 D},0\Bigg\},\nn
		\boldsymbol{v}_{17}&=\Bigg\{\frac{D^4-8 D^2+6}{D^6-13 D^4+36 D^2},\frac{3-D^2}{D^2 \left(D^2-9\right)},-\frac{D (D+3)}{2
			\left(D^4+5 D^3-20 D-16\right)},-\frac{(D-3) D}{2 \left(D^4-5 D^3+20 D-16\right)},\nn
		&\nl \frac{6-4 D^2}{36
			D^2-9 D^4},\frac{2}{9 (D+1) (D+3)},\frac{2}{9 \left(D^2-4 D+3\right)},\frac{1}{18 (D+3)
			(D+4)},\frac{1}{18 \left(D^2-7 D+12\right)},0\Bigg\},\nn
		\boldsymbol{v}_{18}&=\Bigg\{\frac{1}{9 D-D^3},-\frac{1}{18 D-2 D^3},-\frac{D}{4 \left(D^3+3 D^2-6 D-8\right)},\frac{D}{4
			\left(D^3-3 D^2-6 D+8\right)},-\frac{2}{36 D-9 D^3},\nn&\nl\frac{1}{18 \left(D^2+4 D+3\right)},
		-\frac{1}{18
			\left(D^2-4 D+3\right)},\frac{1}{18 (D+3) (D+4)},-\frac{1}{18 (D-4) (D-3)},0\Bigg\},\nn
		\boldsymbol{v}_{19}&=\Bigg\{\frac{1}{9 D-D^3},\frac{2}{D \left(D^2-9\right)},\frac{1}{(D-2) (D+1) (D+4)},\frac{1}{D^3-3
			D^2-6 D+8},\frac{4}{36 D-9 D^3},\nn
		&\nl\frac{2}{9 (D+1) (D+3)},-\frac{2}{9 \left(D^2-4
			D+3\right)},\frac{1}{18 (D+3) (D+4)},-\frac{1}{18 (D-4) (D-3)},0\Bigg\},\nn
		\boldsymbol{v}_{20}&=\Bigg\{\frac{2 D^2-3}{D^2 \left(D^4-13 D^2+36\right)},\frac{D^2-3}{2 D^2
			\left(D^2-9\right)},\frac{D^2+3 D+4}{-4 D^4-20 D^3+80 D+64},\nn
		&\nl \frac{D^2-3 D+4}{-4 D^4+20 D^3-80
			D+64},\frac{D^2+3}{9 D^2 \left(D^2-4\right)},-\frac{1}{9 (D+1) (D+3)},-\frac{1}{9 \left(D^2-4
			D+3\right)},\nn
		&\nl\frac{1}{18 (D+3) (D+4)},\frac{1}{18 \left(D^2-7 D+12\right)},0\Bigg\},\nn
		\boldsymbol{v}_{21}&=\Bigg\{\frac{D^2+6}{D^6-13 D^4+36 D^2},\frac{3}{D^2 \left(D^2-9\right)},\frac{D}{2 \left(D^4+5 D^3-20
			D-16\right)},\frac{D}{-2 D^4+10 D^3-40 D+32},\nn
		&\nl\frac{D^2-6}{9 D^2 \left(D^2-4\right)},-\frac{1}{9
			(D+1) (D+3)},-\frac{1}{9 \left(D^2-4 D+3\right)},\frac{1}{18 (D+3) (D+4)},\frac{1}{18 \left(D^2-7
			D+12\right)},0\Bigg\},\nn
		\boldsymbol{v}_{22}&=\Bigg\{\frac{2 D^2-3}{D^2 \left(D^4-13 D^2+36\right)},\frac{3}{18 D^2-2 D^4},\frac{3 D+4}{4 \left(D^4+5
			D^3-20 D-16\right)},\frac{4-3 D}{4 \left(D^4-5 D^3+20 D-16\right)},\nn
		&\nl\frac{3-2 D^2}{9 D^2
			\left(D^2-4\right)},\frac{1}{18 \left(D^2+4 D+3\right)},\frac{1}{18 \left(D^2-4
			D+3\right)},\frac{1}{18 (D+3) (D+4)},\frac{1}{18 \left(D^2-7 D+12\right)},0\Bigg\},\nn
		\boldsymbol{v}_{23}&=\Bigg\{-\frac{5}{D^5-13 D^3+36 D},\frac{1}{9 D-D^3},\frac{D}{2 \left(D^4+5 D^3-20
			D-16\right)},\frac{D}{2 \left(D^4-5 D^3+20 D-16\right)},\nn
		&\nl\frac{1}{36 D-9 D^3},-\frac{1}{9 (D+1)
			(D+3)},\frac{1}{9 \left(D^2-4 D+3\right)},\frac{1}{18 (D+3) (D+4)},-\frac{1}{18 (D-4) (D-3)},0\Bigg\},\nn
		\boldsymbol{v}_{24}&=\Bigg\{\frac{3}{9 D-D^3},0,\frac{3}{2 (D-2) (D+1) (D+4)},\frac{3}{2 \left(D^3-3 D^2-6
			D+8\right)},\frac{2}{12 D-3 D^3},0,0,\nn
		&\nl-\frac{1}{6 (D+2) (D+3) (D+4)},-\frac{1}{6 \left(D^3-9 D^2+26
			D-24\right)},\frac{2-D^2}{D^5-5 D^3+4 D}\Bigg\},\nn
		\boldsymbol{v}_{25}&=\Bigg\{\frac{5}{D^4-13 D^2+36},0,\frac{1}{2 \left(D^3+7 D^2+14 D+8\right)},\frac{1}{-2 D^3+14 D^2-28
			D+16},0,0,0,\nn
		&\nl-\frac{1}{6 (D+2) (D+3) (D+4)},\frac{1}{6 (D-4) (D-3) (D-2)},\frac{1}{D^4-5 D^2+4}\Bigg\},\nn
		\boldsymbol{v}_{26}&=\Bigg\{-\frac{15}{D^5-13 D^3+36 D},0,-\frac{3}{D^4+5 D^3-20 D-16},\frac{3}{D^4-5 D^3+20
			D-16},-\frac{1}{12 D-3 D^3},\nn
		&\nl0,0,-\frac{1}{6 (D+2) (D+3) (D+4)},-\frac{1}{6 \left(D^3-9 D^2+26
			D-24\right)},-\frac{2}{D^5-5 D^3+4 D}\Bigg\} \nonumber \ed
	\end{align}
	\bibliographystyle{JHEP}
	\bibliography{refer}

\end{document}